\documentclass[12pt]{article}
\usepackage{graphicx,amssymb,amsmath}

\newcommand{\be}{\begin{equation}}
\newcommand{\ee}{\end{equation}}
\newcommand{\bea}{\begin{eqnarray}}
\newcommand{\eea}{\end{eqnarray}}
\newcommand{\beas}{\begin{eqnarray*}}
\newcommand{\eeas}{\end{eqnarray*}}

\begin{document}
\begin{titlepage}
\begin{center}

{\Large Constructing local bulk observables in interacting AdS/CFT}

\vspace{6mm}

\renewcommand\thefootnote{\mbox{$\fnsymbol{footnote}$}}
Daniel Kabat${}^{1}$\footnote{daniel.kabat@lehman.cuny.edu},
Gilad Lifschytz${}^{2}$\footnote{giladl@research.haifa.ac.il}
and David A.\ Lowe${}^{3}$\footnote{lowe@brown.edu}

\vspace{4mm}

${}^1${\small \sl Department of Physics and Astronomy} \\
{\small \sl Lehman College, CUNY, Bronx NY 10468, USA}

\vspace{2mm}

${}^2${\small \sl Department of Mathematics and Physics} \\
{\small \sl University of Haifa at Oranim, Kiryat Tivon 36006, Israel}

\vspace{2mm}

${}^3${\small \sl Department of Physics} \\
{\small \sl Brown University, Providence RI 02912, USA}

\end{center}

\vspace{8mm}

\noindent
Local operators in the bulk of AdS can be represented as smeared operators in the dual CFT.  We show how to
construct these bulk observables by requiring that the bulk operators
commute at spacelike separation.  This extends our previous work by taking interactions into account.  Large-$N$ factorization plays a key role in the construction.
We show diagrammatically how this procedure is related to bulk Feynman diagrams.

\end{titlepage}
\setcounter{footnote}{0}
\renewcommand\thefootnote{\mbox{\arabic{footnote}}}

\section{Introduction}

A fundamental question in quantum gravity is whether one can define
local observables \cite{Kuchar:1991qf,Isham:1992ms}.  The development
of AdS/CFT \cite{Maldacena:1997re} places this question in a new context.  AdS/CFT makes it
clear that, with asymptotic AdS boundary conditions, the physical
degrees of freedom of quantum gravity are completely encoded in the
dual CFT.  In this setting a complete set of observables is provided
by local operators in the CFT.  But local operators in the CFT
only directly describe excitations near the AdS boundary, so the
fundamental question becomes: is there a way to represent local
observables in the bulk using the CFT?

As a closely related question, one of the most puzzling aspects of the
duality between conformal field theories and gravity
is how bulk locality on distance scales
shorter than the anti-de Sitter radius of curvature can be
recovered.\footnote{By bulk locality we mean the existence of local
observables which are causal, i.e.\ which commute at spacelike
separation.}  At the level of two-point functions there is no
obstacle to constructing local observables in an AdS background
\cite{Bena:1999jv,Hamilton:2005ju,Hamilton:2006az,Hamilton:2006fh,Hamilton:2007wj}.
So to probe further one may consider backgrounds that break conformal
symmetry \cite{Lowe:2009mq}, or consider interactions around
backgrounds with exact conformal symmetry
\cite{Heemskerk:2009pn,Heemskerk:2010ty}.

The original dictionary \cite{Gubser:1998bc,Witten:1998qj} is best
thought of as a mapping from bulk AdS correlators to boundary CFT
correlators, in a limit where the bulk operators approach the
boundary.  In \cite{Hamilton:2005ju,Hamilton:2006az,Hamilton:2006fh,Hamilton:2007wj}
we formulated the inverse map in Lorentzian signature, allowing CFT correlators to
be mapped back to bulk correlators.\footnote{The inverse map was derived in
\cite{Dobrev:1998md} as an equivalence of group representations.  It was also derived in
\cite{Bena:1999jv} following the approach of \cite{Banks:1998dd,Balasubramanian:1998de}.}  We worked in the large $N$ limit,
which meant we mapped the CFT to a free theory in the
bulk.\footnote{The map in the opposite limit, from a free CFT
to higher-spin gravity in the bulk, has been discussed in
\cite{Douglas:2010rc}.}  The large-$N$ limit is
rather simple since it sends the bulk Planck length to zero. Can one
construct local observables beyond this limit?  Certainly the usual
lore is that holography forbids the existence of truly local bulk
operators. However since at zero Planck length one can represent the
creation and annihilation operators of the supergravity fields using
CFT data \cite{Banks:1998dd,Balasubramanian:1998de}, one may expect that at least
in some perturbative scheme one can extend the construction to
subleading orders in $1/N$.

In the present paper we address the issue of defining local bulk
observables from CFT data, at subleading orders in $1/N$, by
generalizing the map in \cite{Hamilton:2005ju,Hamilton:2006az,Hamilton:2006fh,Hamilton:2007wj}.
First we show that simply applying the linear smearing transformation
of \cite{Hamilton:2005ju,Hamilton:2006az,Hamilton:2006fh,Hamilton:2007wj}
to a local operator in the CFT leads to correlators with unwanted
singularities, beyond the expected bulk light-cone
singularities. These unwanted singularities imply that the would-be
bulk observables do not commute at spacelike separation, and are not
local from the bulk point of view once interactions are included.
However, as we will show, corrected bulk observables which do commute
at space-like separation may be constructed by mixing in multi-trace
CFT operators with higher conformal dimensions. The relevance of
higher-dimension primary fields to bulk locality was discussed in
\cite{Heemskerk:2009pn,Heemskerk:2010ty,ElShowk:2011ag}, while the
appearance of double-trace operators in internal lines of Witten diagrams was discussed
in \cite{Liu:1998th,D'Hoker:1999jp}.  The condition that the unwanted
singularities can be canceled yields constraints on the CFT, which appear to be satisfied order-by-order in a $1/N$
expansion.  It is possible the cancellation works for a large class
of large $N$ conformal field theories, in line with the conjecture of
\cite{Heemskerk:2009pn}.

In section \ref{sect:problem} we describe the problem of unwanted
singularities, and in section \ref{sect:solution} we give a proposed
solution.  In section \ref{sect:ads2} we carry out the construction of
local bulk observables in AdS${}_2$ and in section \ref{sect:ads3} we present a similar
construction in AdS${}_3$.  In section \ref{sect:bulk} we show how
these results are compatible with bulk perturbation theory (assuming
that such a perturbative description of the bulk is available).  We
present some extensions and generalizations of our results in section
\ref{sect:extensions} and we conclude in section
\ref{sect:conclusions}.

\section{Breakdown of locality\label{sect:problem}}

In this section we show that the definition of a bulk observable given in \cite{Hamilton:2005ju,Hamilton:2006az,Hamilton:2006fh,Hamilton:2007wj}
captures the correct bulk 2-point function.  However when interactions are taken into account it fails to give bulk observables which commute at spacelike
separation.

For concreteness we consider primary operators ${\cal O}_i$ of dimension $\Delta_i$ in a two-dimensional CFT.  The 2- and 3-point functions of these operators
are fixed by conformal invariance.
\bea
&&
\langle {\cal O}_i(x) {\cal O}_j(0) \rangle = {\delta_{ij} \over (X^2 - T^2)^{\Delta_i}} \label{cfttwopoint} \\
\nonumber&&\left\langle \mathcal{O}_{i}({x}_{i})\mathcal{O}_{j}({x}_{j})\mathcal{O}_{k}({x}_{k})\right\rangle =\frac{c_{ijk}}{|{x}_{i}-{x}_{j}|^{\Delta_{i}+\Delta_{j}-\Delta_{k}}|{x}_{i}-{x}_{k}|^{\Delta_{i}+\Delta_{k}-\Delta_{j}}|{x}_{j}-{x}_{k}|^{\Delta_{j}+\Delta_{k}-\Delta_{i}}}\\
\label{eq:cftthreepoint}
\eea
In previous work \cite{Hamilton:2005ju,Hamilton:2006az,Hamilton:2006fh,Hamilton:2007wj} we
considered a free scalar field $\phi$ in the bulk, dual to an operator ${\cal O}$ of dimension $\Delta$ in the CFT,
and showed that the bulk scalar could be reconstructed from the boundary operator via the linear smearing transformation
\begin{eqnarray}
\label{problem:smear}
\phi(Z,X,T)&=&\int dX'dT'K_{\Delta}(Z,X,T|X',T')\mathcal{O}(X',T') \\
\nonumber
&=&\frac{\Delta-1}{\pi}\int_{Y'^{2}+T'^{2}<Z^{2}}dY'dT'\left(\frac{Z^{2}-Y'^{2}-T'^{2}}{Z}\right)^{\Delta-2}\mathcal{O}(X+iY',T+T')
\end{eqnarray}

Applying this transformation to the first operator on the left-hand side of \eqref{cfttwopoint} generates
the expected correlator between one bulk and one boundary point.
To see this we compute (setting $X = 0$ and $T'=r \cos \theta$, $Y'=r \sin \theta$)
\begin{eqnarray}
&& \left\langle \phi(Z,0,T)\mathcal{O}(0,0)\right\rangle \\
\nonumber
&& =\frac{\Delta-1}{\pi}\int_{Y'^2+T'^2<Z^2}dY'dT' \left(\frac{Z^{2}-Y'^{2}-T'^{2}}{Z}\right)^{\Delta-2}\left\langle \mathcal{O}(iY',T+T')\mathcal{O}(0,0)\right\rangle \\
&& =\frac{\Delta-1}{\pi}\left(-1\right)^{\Delta}\int_{0}^{Z}dr\int_{0}^{2\pi}d\theta\left(\frac{Z^{2}-r^{2}}{Z}\right)^{\Delta-2}\frac{r}{\left(r^{2}+T^{2}+2rT\cos\theta\right)^{\Delta}}
\label{2point}
\end{eqnarray}
For the moment we assume $T>Z$ and later use analytic continuation
to generalize. We use the result
\begin{equation}
\int_{0}^{2\pi}d\theta\frac{1}{\left(r^{2}+T^{2}+2rT\cos\theta\right)^{\Delta}}=2\pi T^{-2\Delta}\,_{2}F_{1}\left(\Delta,\Delta;1;\frac{r^{2}}{T^{2}}\right)\label{eq:thetaintegral}\end{equation}
to perform the $\theta$ integral.  The $r$ integral then becomes,
defining $q=r^{2}/Z^{2}$ and $y=Z^{2}/T^{2}$,
\begin{eqnarray*}
\left\langle \phi(Z,0,T)\mathcal{O}(0,0)\right\rangle  & = & \frac{\Delta-1}{2\pi R}\left(-1\right)^{\Delta}T^{-2\Delta}Z^{\Delta}\int_{0}^{1}dq(1-q)^{\Delta-2}\,_{2}F_{1}\left(\Delta,\Delta;1;qy\right)\\
 & = & \frac{Z^{\Delta}}{2\pi R}\frac{1}{\left(Z^{2}-T^{2}\right)^{\Delta}}\,.\end{eqnarray*}
The result for general $X,T$ may be obtained using Lorentz
invariance and analytic continuation.  With a Wightman $i \epsilon$ prescription
\begin{equation}
\left\langle \phi(Z,X,T)\mathcal{O}(0,0)\right\rangle =\frac{Z^{\Delta}}{2\pi R}\,\frac{1}{\left(Z^{2}+X^{2}-(T-i\epsilon)^{2}\right)^{\Delta}}
\label{2pointresult}
\end{equation}
This is the expected bulk-boundary two-point function for AdS${}_3$ in Poincar\'e coordinates.  Note that inside the two point
function the operators will commute at bulk space-like separation.
So at this level bulk locality appears to be compatible with the definition (\ref{problem:smear})
of a bulk observable.  As we now show, interactions change this conclusion.

To take interactions into account we study a three point function.  For simplicity we consider three operators of dimension $\Delta=2$.  Up to an overall
coefficient, their three point function reads
\be
\langle\mathcal{O}(x_0) \mathcal{O}(x_1)\mathcal{O}(x_2)\rangle ={1 \over \vert x_0 - x_1 \vert^2 \vert x_0 - x_2 \vert^2 \vert x_1 - x_2 \vert^2}
\ee
We now smear the first operator using (\ref{problem:smear}), to turn it into a bulk
operator.  We need to do the integral (note that the last term on the right hand side
just comes along for the ride, so will be dropped)
\begin{eqnarray}
&&\langle\phi(Z,X_0,T_0) \mathcal{O}(X_1,T_1)\mathcal(X_2,T_2)\rangle \\
\nonumber
&&\sim\int_{Y'^2+T'^2<Z^2}\frac{dY'dT'} {[(T_{1}-T_{0}-T')^2-(X_{1}-X_{0}-iY')^{2}][(T_{2}-T_{0}-T')^2-(X_{2}-X_{0}-iY')^{2}]}
\end{eqnarray}
Defining $T'=r \cos \theta$ and $Y'=r \sin \theta$ with $\alpha =e^{i\theta}$, and denoting
$X^{+}_{kl}=(T+X)_{k}-(T+X)_{l}$ and $X^{-}_{kl}=(T-X)_{k}-(T-X)_{l}$, we get
\begin{equation}
\int_{0}^{Z}rdr\int \alpha d\alpha[(X^{+}_{10}-r\alpha)(\alpha X^{-}_{10}-r)(X^{+}_{20}-r\alpha)(\alpha X^{-}_{20}-r)]^{-1}
\end{equation}
where the integral over $\alpha$ is a contour integral around the poles at $\alpha=\frac{r}{X^{-}_{10}}$ and $\alpha=\frac{r}{X^{-}_{20}}$.
Doing the integrals gives
\begin{equation}
\label{eq:nonlocalthree}
\frac{1}{(X_1 -X_2)^2-(T_1 -T_2)^2}\left[\ln \frac{Z^2 -X^{+}_{10}X^{-}_{10}}{Z^2 -X^{+}_{20}X^{-}_{10}} +
\ln \frac{Z^2 -X^{+}_{20}X^{-}_{20}}{Z^2 -X^{+}_{10}X^{-}_{20}}\right]
\end{equation}
This result is AdS covariant, as we show in appendix
\ref{sect:boundary-to-bulk-map}.  However the terms $\ln \big(Z^2
-X^{+}_{20}X^{-}_{10}\big)$ and $\ln \big(Z^2
-X^{+}_{10}X^{-}_{20}\big)$ give rise to singularities (and hence
non-zero commutators) even when all three operators are spacelike
separated.  This means the prescription (\ref{problem:smear}) for
defining bulk operators in the CFT cannot be used beyond the leading
large-$N$ limit (that is, when the bulk theory is not free).

Another way to reach the same conclusion is to study the OPE between quasi-primary operators.
\begin{equation}
\mathcal{O}_{i}(X,T)\mathcal{O}_{j}(0)=\frac{\delta_{ij}}{(X^{2}-T^{2})^{\Delta_{i}}}+\sum_{k}\frac{c_{ijk}}{(X^{2}-T^{2})^{(\Delta_{i}+\Delta_{j}-\Delta_{k})/2}}\mathcal{O}_{k}(0)+\cdots \label{eq:operel}\end{equation}
For simplicity we specialize to a dimension two operator with
\begin{equation}
\label{BdyOPE}
{\cal{O}} (X,T){\cal{O}} (0)=\frac{1}{(X^2-T^2)^2}+\frac{1}{N} \frac{{\cal{O}}(0)}{(X^2-T^2)} + \cdots
\end{equation}
(the $1/N$ coefficient reflects large-$N$ counting).  Let's try to use the smearing
transformation (\ref{problem:smear}) to turn this into a bulk - boundary OPE.
The first term in (\ref{BdyOPE}) just gives the bulk-boundary two-point function, but the second term gives
\begin{equation}
\frac{{\cal{O}}(0)}{\pi N}\int_{Y'^{2}+T'^{2}<Z^2}\frac{1}{(X+iY')^2 - (T+T')^2}
\end{equation}
Unlike the 3-point correlator considered above this
integral is not AdS covariant.\footnote{Since the OPE is a short-distance expansion in the CFT it does not lift to a covariant OPE in the bulk.}
We can do the integral by going to $r,\alpha$ variables as before, and we get 
\begin{eqnarray}
&&\sim \frac{{\cal{O}}(0)}{N}\int_{0}^{Z}rdr\oint_{|\alpha|=1}d\alpha\frac{1}{(T+X+r\alpha)(\alpha(T-X)+r)}\nonumber\\
\label{problem:OPE}
&&= \frac{{\cal{O}}(0)}{2N}\,\ln \frac{X^2 - T^2}{X^2 + Z^2 -T^2}
\end{eqnarray}
Besides the expected bulk light-cone singularity, at $X^2 + Z^2 = T^2$, there is a boundary
light-cone singularity at $X^2 = T^2$.  Again these unwanted singularities (which are
not even AdS covariant) mean that operators will not commute at bulk space-like separation.

This means the boundary-to-bulk map constructed in
\cite{Hamilton:2005ju,Hamilton:2006az,Hamilton:2006fh,Hamilton:2007wj},
if applied to an interacting CFT, gives rise to a set of bulk
observables which are non-local in the sense that they do not commute
at spacelike separation.  These non-local observables could still be
used to study bulk physics.  However it's natural to ask if there is a
way of constructing bulk observables in an interacting CFT.  This is
the question we address in the remainder of the paper.

\section{A possible cure\label{sect:solution}}

Since we still want a bulk scalar field, there are only a limited
number of ways of changing the original construction
(\ref{problem:smear}). Given an operator of dimension $\Delta$ the
smearing function we used is the unique way of mapping it to a bulk
scalar field. So the only possible deformation of our construction is
to add higher dimension, appropriately smeared primary operators
(assuming such operators are available).  With a sum over CFT primaries, the definition of a bulk operator becomes
\begin{equation}
\phi(Z,X,T)=\int K_{\Delta}(Z,X,T|X',T')\mathcal{O}(X',T')+\sum_{k}d_{k}\int K_{\Delta_{k}}(Z,X,T|X',T')\mathcal{O}_{\Delta_{k}}(X',T')\,.
\label{eq:bulkdress}
\end{equation}
Here $K_{\Delta_{k}}$ is the appropriate AdS covariant smearing
function for an operator of dimension $\Delta_{k}$.  As we will see
later the terms we have added produce $\log$ singularities of the
type we found in the previous section, times a polynomial in $(X^{2}-T^{2})/Z^{2}$. The
coefficients $d_{k}$ can be fixed (or at least constrained) by
demanding that the unwanted $\log$ singularities appearing in
(\ref{eq:nonlocalthree}), (\ref{problem:OPE}) are canceled to some
order in $(X^{2}-T^{2})/Z^{2}$ (or perhaps to all orders).  Of
course this cancellation requires the existence of primary fields with
increasing dimensions, with appropriate OPE's.  If such operators are
unavailable then bulk locality is destroyed on macroscopic scales.

The two-point function one recovers from this procedure is consistent
with the general form of a two-point function one would expect based
on a spectral decomposition
\begin{equation}
\left\langle \phi(Z,X,T)\phi(Z',X',T')\right\rangle _{bulk}=\int dm^{2}\rho(m^{2})G_{0}(Z,X,T|Z',X',T';m^{2})\label{eq:lehmann}
\end{equation}
where $G_{0}(.|.;m^{2})$ is the free two-point function for a scalar
field of mass $m^{2}$ and $\rho(m^{2})$ is the positive semi-definite
spectral density. It is worth making a few remarks on the formula
\eqref{eq:lehmann} that are somewhat surprising from the viewpoint
of flat space quantum field theory. In general, the bounds of the
integral range over all possible values of the mass allowed by unitarity.
However this is puzzling from the CFT viewpoint, since it appears
to require a continuous spectrum of quasi-primary operators. Typically well-behaved conformal field theories have
a discrete tower of primary operators. The puzzle is resolved once
one realizes that at least in bulk perturbation theory, the density
of states $\rho(m^{2})$ is typically not a continuous function. Rather
the AdS symmetries pick out discrete towers of masses that
arise when, for example, an interaction polynomial in a scalar field
is expanded in perturbation theory \cite{Dusedau:1985ue}. Thus, for
example, a scalar field $\phi$ with mass $m$ and interaction $\lambda\phi^{3}$
would give rise to terms in \eqref{eq:lehmann} dual to CFT operators
of conformal weight $\Delta+n$, as well as weights $2\Delta+n$,
$3\Delta+n$, $\cdots$, with $n$ a non-negative integer.

In the remainder of this paper we show in explicit examples that, at least in CFT's with a
$1/N$ expansion, it seems possible to construct the higher-dimension operators which
are necessary for bulk locality, as multi-trace operators with derivatives.

\section{CFT construction: AdS${}_2$\label{sect:ads2}}

In this section we show that one can correct the definition of a bulk
observable in such a way as to restore bulk locality.  For simplicity
we begin with AdS${}_2$; in the next section we treat AdS${}_3$.

As a guide, in section \ref{sect:correlators} we review
correlators in AdS${}_2$ / CFT${}_1$.  In section \ref{sect:lowest} we
apply the linear smearing transformation to 2- and 3-point functions in the CFT
and show that the resulting bulk operators fail to commute at
spacelike separation.
In section \ref{sect:HDO} we argue that this
can be cured by adding an infinite sequence of higher-dimension
operators; we construct the necessary operators using a $1/N$ expansion.
In section \ref{sect:bilocal} we show that another way to restore
spacelike commutativity at ${\cal O}(1/N)$ is to add a bilocal correction term.
This bilocal correction can be thought of as resumming the tower of
higher-dimension operators.

\subsection{AdS correlators\label{sect:correlators}}

We work in the Poincar\'e patch of AdS${}_2$ with metric
\[
ds^2 = {R^2 \over Z^2} \left(-dT^2 + dZ^2\right)
\]
We consider a massless scalar field $\phi$, dual to a dimension-1 operator ${\cal O}$ in the CFT.
That is, as $Z \rightarrow 0$ we have
\[
\phi(T,Z) \rightarrow Z \, {\cal O}(T)\,.
\]
The free bulk two-point function is \cite{Spradlin:1999bn}
\be
\label{Bulk2pt}
\langle \phi(T,Z) \phi(T',Z') \rangle = {1 \over 2\pi} \tanh^{-1} \big(1/\sigma\big)
\ee
where the invariant distance
\be
\label{InvDistance}
\sigma = {Z^2 + Z'{}^2 - (T - T')^2 \over 2 Z Z'}\,.
\ee
Sending one point to the boundary gives the mixed bulk-boundary correlator
\be
\label{BulkBdy}
\langle \phi(T,Z) {\cal O}(T') \rangle = {1 \over \pi} \, {Z \over Z^2 - (T - T')^2}
\ee
while sending both points to the boundary gives the CFT correlator
\be
\label{Bdy2pt}
\langle {\cal O}(T) {\cal O}(T') \rangle = - {1 \over \pi} \, {1 \over (T - T')^2}\,.
\ee

So far we haven't given a prescription for handling light-cone
singularities.  The correct prescription depends on which Green's
function you want.  The Wightman function is defined by $T \rightarrow
T - i \epsilon$, while the Feynman function is defined by $(T - T')^2
\rightarrow (T - T')^2 - i \epsilon$.  So for instance
\bea
\label{Wightman}
&& \langle 0 \vert {\cal O}(T) {\cal O}(T') \vert 0 \rangle = - {1 \over \pi} \, {1 \over (T - T' - i \epsilon)^2} \\
\label{Feynman}
&& \langle 0 \vert T\lbrace{\cal O}(T) {\cal O}(T')\rbrace \vert 0 \rangle = - {1 \over \pi} \, {1 \over (T - T')^2 - i \epsilon}
\eea

We'll also need the 3-point correlator in the CFT.  Provided that $T_1 > T_2 > T_3$ this is given by
\be
\label{Bdy3pt}
\langle 0 \vert {\cal O}(T_1) {\cal O}(T_2) {\cal O}(T_3) \vert 0 \rangle = - {i \lambda R^2 \over \pi} \,
{1 \over (T_1 - T_2) (T_1 - T_3) (T_2 - T_3)}
\ee
Here $\lambda R^2$ is a dimensionless coefficient.  As we'll discuss in appendix \ref{sect:3point}
this is induced at tree level by a bulk $\lambda \phi^3$ interaction.  However aside from the
coefficient the form of this result is fixed by conformal invariance.  It can
be continued outside the range $T_1 > T_2 > T_3$ with suitable $i
\epsilon$ prescriptions.  For instance suppose we wanted to extend
(\ref{Bdy3pt}) past the singularity at $T_1 = T_2$ without changing
the operator ordering.
This can be done with a $T_1 \rightarrow T_1 - i \epsilon$
prescription:
\be
\label{Bdy3pt2}
\langle 0 \vert {\cal O}(T_1) {\cal O}(T_2) {\cal O}(T_3) \vert 0 \rangle = - {i \lambda R^2 \over \pi} \,
{1 \over (T_1 - T_2 - i \epsilon) (T_1 - T_3) (T_2 - T_3)}
\ee
This is the same prescription used to handle singularities in the Wightman function (\ref{Wightman}).
It can be understood as a way to regulate the time evolution operator $e^{-i H (T_1 - T_2)}$.  Other choices
are possible, for instance the time-ordered 3-point function is given in (\ref{Feynman3pt}).

\subsection{Linear smearing\label{sect:lowest}}

At lowest order we have the linear smearing relation \cite{Hamilton:2005ju}
\be
\label{LowestOrder}
\phi^{(0)}(T,Z) = {1 \over 2} \int_{T-Z}^{T+Z} dT_1 \, {\cal O}(T_1)\,.
\ee
In this section we use this relation to generate candidate bulk observables.  We'll show that everything works fine
at the level of 2-point functions.  But when we consider 3-point functions we'll see that the bulk operators we
construct fail to commute at spacelike separation.

To illustrate the procedure, consider smearing one leg of the CFT 2-point function (\ref{Wightman}).
This should give a mixed bulk - boundary correlator.  Using a Wightman $i \epsilon$ prescription we find
\bea
\nonumber
\langle 0 \vert \phi^{(0)}(T,Z) {\cal O}(T') \vert 0 \rangle & = &
{1 \over 2} \int_{T-Z}^{T+Z} dT_1 \, \left(- {1 \over \pi}\right) {1 \over \big(T_1 - T' - i \epsilon \big)^2} \\
\label{Lowest2pt}
& = & {1 \over \pi} \, {Z \over Z^2 - (T - T' - i \epsilon)^2}
\eea
which reproduces the exact result (\ref{BulkBdy}).  Likewise smearing the second leg gives
\bea
\nonumber
\langle 0 \vert \phi^{(0)}(T,Z) \phi^{(0)}(T',Z') \vert 0 \rangle &=&
{1 \over 2} \int_{T'-Z'}^{T'+Z'} dT'_1 \,\, {1 \over \pi} {Z \over Z^2 - (T - T'_1 - i \epsilon)^2} \\
&=& {1 \over 2\pi} \tanh^{-1} \left({2 Z Z' \over Z^2 + Z'{}^2 - (T - T' - i \epsilon)^2}\right) \qquad
\eea
in agreement with the bulk Wightman function (\ref{Bulk2pt}).  The $i \epsilon$
prescriptions here cause no difficulty: smearing the Wightman function in the CFT gives the correct
bulk Wightman function.  As we will see in section \ref{sect:Feynman}, the story is more complicated
for Feynman propagators.

So far, so good.  But now let's see what happens when we apply the linear smearing relation (\ref{LowestOrder}) to the
first operator in the CFT 3-point function (\ref{Bdy3pt}).
Taking $T - Z > T_2 > T_3$ so that we don't need to worry about $i \epsilon$ prescriptions, the integral gives
\be
\label{phi0OO}
\langle 0 \vert \phi^{(0)}(T,Z) {\cal O}(T_2) {\cal O}(T_3) \vert 0 \rangle = {i \lambda R^2 \over 2 \pi} \,
{1 \over (T_2 - T_3)^2} \, \log {(T + Z - T_3) (T - Z - T_2) \over (T + Z - T_2) (T - Z - T_3)}\,.
\ee
This result has some nice properties.  It only has singularities when the bulk point is lightlike-separated from
one of the boundary points.  Also it's covariant under
$SO(1,2)$.\footnote{The prefactor $1/(T_2 - T_3)^2$ has the right conformal
weight, and you can check that the argument of the logarithm is
invariant under the special conformal transformation
\begin{eqnarray*}
T & \rightarrow & {T + b (T^2 - Z^2) \over 1 + 2 b T + b^2 (T^2 - Z^2)} \\
Z & \rightarrow & {Z \over 1 + 2 b T + b^2 (T^2 - Z^2)}
\end{eqnarray*}
The boundary points transform as $T_2 \rightarrow T_2 / (1 + b
T_2)$, $T_3 \rightarrow T_3 / (1 + b T_3)$.}

Despite these nice properties, the bulk operators we have constructed
don't commute at spacelike separation.  To see this we first continue
(\ref{phi0OO}) into the regime $T+Z > T_2 > T-Z > T_3$, using a $T_2
\rightarrow T_2 + i \epsilon$ prescription to avoid the singularity at
$T_2 = T - Z$.  This gives
\be
\label{phi0O}
\langle 0 \vert \phi^{(0)}(T,Z) {\cal O}(T_2) {\cal O}(T_3) \vert 0 \rangle = {i \lambda R^2 \over 2 \pi} \,
{1 \over (T_2 - T_3)^2} \, \log {(T + Z - T_3) (T - Z - T_2 - i \epsilon) \over (T + Z - T_2) (T - Z - T_3)}\,.
\ee
Then we repeat the calculation, starting from
\be
\langle 0 \vert {\cal O}(T_2) {\cal O}(T_1) {\cal O}(T_3) \vert 0 \rangle = + {i \lambda R^2 \over \pi} \,
{1 \over (T_1 - T_2) (T_1 - T_3) (T_2 - T_3)}
\ee
which is valid for $T_2 > T_1 > T_3$.  Note the change of sign!  Smearing the middle operator and continuing
to $T+Z > T_2 > T-Z > T_3$ with a $T_2 \rightarrow T_2 - i \epsilon$ prescription gives
\be
\label{Ophi0}
\langle 0 \vert {\cal O}(T_2) \phi^{(0)}(T,Z) {\cal O}(T_3) \vert 0 \rangle = - {i \lambda R^2 \over 2 \pi} \,
{1 \over (T_2 - T_3)^2} \, \log {(T + Z - T_3) (T_2 - T + Z) \over (T_2 - T - Z - i \epsilon) (T - Z - T_3)}\,.
\ee
Taking the difference of (\ref{phi0O}) and (\ref{Ophi0}) gives the commutator
\be
\label{LowestCommutator}
\langle 0 \vert i \big[\phi^{(0)}(T,Z), {\cal O}(T_2)\big] {\cal O}(T_3) \vert 0 \rangle = - {\lambda R^2 \over \pi} \,
{1 \over (T_2 - T_3)^2} \, \log {(T + Z - T_3) (T_2 - T + Z) \over (T + Z - T_2) (T - Z - T_3)}\,.
\ee
This is non-vanishing at spacelike separation.

\subsection{Higher dimension operators\label{sect:HDO}}

Let's see if we can add something to the lowest-order bulk operator (\ref{LowestOrder}) that will
restore spacelike commutativity.  The only objects at our disposal would seem to be higher-dimension
operators.  For instance at large $N$ we can build a dimension-2 primary field\footnote{The colons
denote normal-ordering, {\em i.e.}~no self-contractions.  The statement that ${\cal O}_2$
has dimension 2 is true at large $N$, where
we can ignore anomalous dimensions and operator mixing.}
\[
{\cal O}_2(T) = \colon{\cal O}(T) {\cal O}(T)\colon
\]
and we could imagine adding a correction term
\be
\label{phi12}
\phi^{(1)}_{\Delta = 2}(T,Z) = A \int_{T-Z}^{T+Z} dT' \, {Z^2 - (T-T')^2 \over Z} \, {\cal O}_2(T')\,.
\ee
Here $A$ is a coefficient we need to determine, and we've used the smearing
function $\sim (\sigma Z')^{\Delta - 1}$ appropriate to a dimension-2 operator.  Likewise at dimension 4 we have
a primary field
\[
{\cal O}_4(T) = \colon\partial_T {\cal O} \partial_T {\cal O} - {2 \over 3} O \partial_T^2 O\colon
\]
and we could imagine adding a correction
\[
\phi^{(1)}_{\Delta = 4}(T,Z) = B \int_{T-Z}^{T+Z} dT' \left({Z^2 - (T-T')^2 \over Z}\right)^3 {\cal O}_4(T')\,.
\]

In this way we have an infinite number of parameters $A,B,\ldots$ at our disposal.  The idea is to fix these
coefficients so as to cancel off the commutator (\ref{LowestCommutator}).  It's useful to work in terms
of
\be
\label{psi}
\psi = {Z^2 - T^2 + T T_2 + T T_3 - T_2 T_3 \over Z(T_2 - T_3)}\,.
\ee
This is the unique $SO(2,1)$-invariant quantity associated with one bulk point $(T,Z)$ and two boundary points
$T_2$, $T_3$.  The regime of interest, where the bulk point is spacelike separated from the first boundary point,
corresponds to $T-Z < T_2 < T+Z$ or equivalently $-1 < \psi < 1$.

The lowest-order commutator calculated in (\ref{LowestCommutator}) can be expressed in terms of $\psi$.
\beas
\langle 0 \vert i \big[\phi^{(0)}(T,Z), {\cal O}(T_2)\big] {\cal O}(T_3) \vert 0 \rangle
&=& - {\lambda R^2 \over \pi} \, {1 \over (T_2 - T_3)^2} \, \log {1 + \psi \over 1 - \psi} \\
&=& - {2 \lambda R^2 \over \pi} \, {1 \over (T_2 - T_3)^2} \left(\psi + {1 \over 3} \psi^3 + {1 \over 5} \psi^5 + \cdots \right) \,.
\eeas
In appendix \ref{appendix:HDO} we show that at leading order for large $N$
\beas
\langle 0 \vert i \big[\phi^{(1)}_{\Delta = 2}(T,Z), {\cal O}(T_2)\big] {\cal O}(T_3) \vert 0 \rangle
&=& {8 A \over \pi} \, {1 \over (T_2 - T_3)^2} \, \psi \\
\langle 0 \vert i \big[\phi^{(1)}_{\Delta = 4}(T,Z), {\cal O}(T_2)\big] {\cal O}(T_3) \vert 0 \rangle
&=& {96 B \over \pi} \, {1 \over (T_2 - T_3)^2} \Big(\psi - {5 \over 3} \psi^3\Big)
\eeas
These results rely on large-$N$ factorization: they were obtained from a disconnected product of 2-point
functions in the CFT, which makes the leading contribution at large $N$.  The
connected 4-point correlator of the CFT, which is a subleading correction
in the $1/N$ expansion, would modify these results.

Using just the dimension-2 primary we could cancel the term linear in $\psi$ by
setting $A = {1 \over 4} \lambda R^2$.  Using both dimension-2 and dimension-4 primaries we could cancel
the $\psi$ and $\psi^3$ terms by setting $A = {3 \over 10} \lambda R^2$, $B = - {1 \over 240} \lambda R^2$.
Assuming this pattern holds in general, by including operators up to dimension $\Delta$ we could cancel
the first $\Delta/2$ terms in the Taylor series expansion of the commutator.

As we'll see in the next section, it's possible to re-sum this infinite series to
obtain a correction term which is bilocal in ${\cal O}(T)$.  These results will show that the series converges, with
$A \rightarrow {3 \over 8} \lambda R^2$ as more and more operators are taken into account.

\subsection{Bilinear smearing\label{sect:bilocal}}

It's not hard to write down a correction to the lowest order smearing function (\ref{LowestOrder}) which fully
restores spacelike commutativity at ${\cal O}(1/N)$.  Consider the bilocal operator
\be
\label{FirstOrder}
\phi^{(1)}(T,Z) = {\lambda R^2 \over 8} \int_0^Z {dZ' \over Z'{}^2} \int_{T - Z + Z'}^{T + Z - Z'} dT'
\int_{T'-Z'}^{T'+Z'} dT_1 dT_2 \, \colon {\cal O}(T_1) {\cal O}(T_2) \colon
\ee
Here $\colon \, \cdots \, \colon$ denotes normal-ordering (meaning no
self-contractions).  The $(T',Z')$ integrals run over the right
light-cone of the bulk point.  The claim is that, if one ignores 4-
and higher-point functions in the CFT, the operator $\phi^{(0)} +
\phi^{(1)}$ commutes at spacelike separation.  In the $1/N$ expansion, this corresponds to ignoring ${\cal O}(1/N^2)$
effects.\footnote{At this order in $1/N$ the operator ordering doesn't matter.  But the results
of section \ref{sect:Feynman} suggest that it's natural to time-order the operators appearing
on the right-hand side of (\ref{FirstOrder}).}

To show that adding $\phi^{(1)}$ makes the commutator vanish we first take $T - Z > T_2 > T_3$, where
\beas
&& \langle 0 \vert \phi^{(1)}(T,Z) {\cal O}(T_2) {\cal O}(T_3) \vert 0 \rangle = 
{\lambda R^2 \over 8} \int_0^Z {dZ' \over Z'{}^2} \int_{T - Z + Z'}^{T + Z - Z'} dT'
\int_{T'-Z'}^{T'+Z'} d\tilde{T}_1 d\tilde{T}_2 \\
&& \qquad \langle 0 \vert \colon {\cal O}(\tilde{T}_1) {\cal O}(\tilde{T}_2) \colon
{\cal O}(T_2) {\cal O}(T_3) \vert 0 \rangle
\eeas
At this stage we have to evaluate a 4-point correlator in the CFT.  Again we use large-$N$ factorization,
which tells us that at leading order for large $N$ the correlator is given by a disconnected product of CFT 2-point functions.
This approximation gives
\bea
\label{phi1O}
&& \langle 0 \vert \phi^{(1)}(T,Z) {\cal O}(T_2) {\cal O}(T_3) \vert 0 \rangle = 
{\lambda R^2 \over \pi^2} \int_0^Z dZ' \int_{T - Z + Z'}^{T + Z - Z'} dT' \, \\
\nonumber
&& \qquad {1 \over (T' + Z' - T_2)(T' + Z' - T_3)(T' - Z' - T_2)(T' - Z' - T_3)}
\eea
Of course taking the connected 4-point correlator of the CFT into account, which is a subleading effect in the $1/N$
expansion, would change this result.

The next step is to continue (\ref{phi1O}) into the regime $T + Z > T_2 > T - Z > T_3$
using a $T_2 \rightarrow T_2 + i \epsilon$ prescription.  A similar calculation of
$\langle 0 \vert {\cal O}(T_2) \phi^{(1)}(T,Z) {\cal O}(T_3) \vert 0 \rangle$ leads to
exactly the same expression, but with a $T_2 \rightarrow T_2 - i \epsilon$
prescription.  Taking the difference, the commutator is given by
integrating $T'$ over a closed contour.  The contour encircles the
pole at $T' = T_2 + Z'$ provided $0 < Z' < (T + Z - T_2) / 2$, and it
encircles the pole at $T' = T_2 - Z'$ provided $0 < Z' < (T_2 - T + Z) / 2$.  So
\bea
\nonumber
&& \langle 0 \vert i \big[\phi^{(1)}(T,Z), {\cal O}(T_2)\big] {\cal O}(T_3) \vert 0 \rangle \\
\nonumber
&& = - {2 \lambda R^2 \over \pi} \, {1 \over (T_2 - T_3)} \left[ \int_0^{(T + Z - T_2)/2} {dZ' \over 2 Z' (2 Z' + T_2 - T_3)}
+ \int_0^{(T_2 - T + Z)/2} {dZ' \over 2 Z' (2 Z' + T_3 - T_2)} \right] \\
&& = {\lambda R^2 \over \pi} \, {1 \over (T_2 - T_3)^2} \,
\log {(T + Z - T_3) (T_2 - T + Z) \over (T + Z - T_2) (T - Z - T_3)}
\eea
This exactly cancels (\ref{LowestCommutator}).

To make contact with the results of the previous section, consider expanding (\ref{FirstOrder}) in powers of $Z$.
Near the boundary the leading behavior is
\[
\phi^{(1)}(T,Z) \sim {1 \over 2} \lambda R^2 Z^2 \, \colon \big({\cal O}(T)\big)^2 \colon \quad {\rm as} \quad
Z \rightarrow 0
\]
The interpretation is that we've corrected the lowest-order smearing
function (\ref{LowestOrder}) by mixing in a dimension-2 operator.
Matching to the behavior of (\ref{phi12}) near the boundary, namely
\[
\phi^{(1)}_{\Delta = 2}(T,Z) \sim {4 \over 3} A Z^2 \, \colon \big({\cal O}(T)\big)^2 \colon
\]
fixes $A = {3 \over 8} \lambda R^2$.  Subleading terms in the expansion of $\phi^{(1)}$ correspond to the infinite
sequence of higher dimension operators considered in the previous section.

\section{CFT construction: AdS$_{3}$\label{sect:ads3}}

We now consider the construction of bulk observables in AdS$_{3}$. As we showed in section
\ref{sect:problem}, once interactions are taken into account the bulk observables defined in
\cite{Hamilton:2005ju,Hamilton:2006az,Hamilton:2006fh,Hamilton:2007wj}
do not commute with boundary operators, even when
the bulk and boundary points are at spacelike separation.  As in section \ref{sect:HDO}
we will cure this problem by adding higher dimension operators to our definition of a bulk observable.
Our conclusions in this section are based on smearing the OPE in the CFT.  Analogous results, based on smearing CFT correlators,
are obtained in appendix \ref{sect:General-Mixed-Bulk/Boundary}.

Imagine we have an infinite set of primary operators ${\cal O}_{i}$ with
dimension $\Delta_{i}$, with OPE
\be
\label{AdS3CFTOPE}
{\cal O}_{{i}}(X,T){\cal O}_{{j}}(0,0)= {\delta_{ij} \over (X^2 - T^2)^{\Delta_i}} + c_{ijk}\frac{{\cal O}_{{k}}(0,0)}
{(X^2-T^2)^{\tilde{\Delta}}}+ \cdots
\ee
Here $\tilde{\Delta}=(\Delta_{i}+\Delta_{j}-\Delta_{k})/2$. 
Using (\ref{problem:smear}) we smear ${\cal O}_{i}$ to turn it into a bulk operator
$\phi_{{i}}(Z,X,T)$.  The first term in the OPE gives the free bulk - boundary 2-point function, while the second gives
\be
\label{AdS3BulkOPE}
\phi_{{i}}(Z,X,T) {\cal O}_{{j}}(0,0)=c_{ijk}f(Z,X,T;\,0,0){\cal O}_{{k}}(0,0)+\cdots
\ee
where
\beas
&& f(Z,X,T;\,0,0) \\
&& = \frac{\Delta_{i}-1}{\pi}\left(-1\right)^{\tilde{\Delta}}\int_{Y'^2+T'^2<Z^2}dY'dT' \left(\frac{Z^{2}-Y'^2-T'^2}{Z}\right)^{\Delta_{i}-2}\frac{1}{\left((T+T')^2-(X+iY')^2 \right)^{\tilde{\Delta}}}
\eeas
As before we begin by working in the regime $T>Z$ with $X=0$.  Switching to polar coordinates
\[
f(Z,0,T;\,0,0)=\frac{\Delta_{i}-1}{\pi}\left(-1\right)^{\tilde{\Delta}}\int_{0}^{Z}dr\int_{0}^{2\pi}d\theta\left(\frac{Z^{2}-r^{2}}{Z}\right)^{\Delta_{i}-2}\frac{r}{\left(r^{2}+T^{2}+2rT\cos\theta\right)^{\tilde{\Delta}}}
\]
Compared to the two point function (\ref{2point}), the only difference is the relative power of the two factors in the
integrand.  The integral in (\ref{2point}) reflected the casual structure of AdS and only had singularities on the bulk
lightcones.  Here things will be different.

The integral over $\theta$ is performed as before using \eqref{eq:thetaintegral}.
Again defining $q=r^{2}/Z^{2}$ and $y=Z^{2}/T^{2}$ we obtain
\begin{eqnarray*}
f(Z,0,T;0,0) & = & \left(\Delta_{i}-1\right)\left(-1\right)^{\tilde{\Delta}}T^{-2\tilde{\Delta}}Z^{\Delta_{i}}\int_{0}^{1}dq(1-q)^{\Delta_{i}-2}\,_{2}F_{1}\left(\tilde{\Delta},\tilde{\Delta};1;qy\right)\\
& = & \left(-1\right)^{\tilde{\Delta}}T^{-2\tilde{\Delta}}Z^{\Delta_{i}}\,_{2}F_{1}\left(\tilde{\Delta},\tilde{\Delta};\Delta_{i};\frac{Z^{2}}{T^{2}}\right)\,.\end{eqnarray*}
We can extend this to general $X,T$ using analytic continuation
and Lorentz invariance.
\begin{equation}
f(Z,X,T;0,0)=\left(X^{2}-\left(T-i\epsilon\right)^{2}\right)^{-\tilde{\Delta}}Z^{\Delta_{i}}\,_{2}F_{1}\left(\tilde{\Delta},\tilde{\Delta};\Delta_{i};\frac{Z^{2}}{(T-i\epsilon)^{2}-X^{2}}\right)\,.\label{eq:threeptcoeff}\end{equation}

Let's look at a few relevant limits of this expression.  First, in the limit $Z \to 0$ with $X,T$ fixed,
we have $\phi(Z,X,T)\to Z^{\Delta_{i}}\mathcal{O}(X,T)$
by construction \cite{Hamilton:2006az}. In this limit\begin{equation}
f(Z,X,T;0,0)=\left(X^{2}-\left(T-i\epsilon\right)^{2}\right)^{-\tilde{\Delta}}Z^{\Delta_{i}}\label{eq:threepointleading}\end{equation}
So indeed in this limit the mixed bulk-boundary OPE (\ref{AdS3BulkOPE}) goes over to the CFT OPE (\ref{AdS3CFTOPE}).

To see the failure of bulk locality we need to look at a different limit where we approach a boundary lightcone.
Let's first look at the case where all operators have even dimensions. Then the hypergeometric function has a simple
form in terms of elementary functions,
\be
{}_{2}F_{1}\left(\tilde{\Delta},\tilde{\Delta};\Delta_{i};z\right)\,=
R_{1}(z)+R_{2}\ln (1-z)
\ee
where $R_{i}(z)$ are rational functions of $z=\frac{Z^{2}}{(T-i\epsilon)^{2}-X^{2}}$. In fact one can show that 
\be
f(Z,X,T;0,0)= g_{1}(z)+Z^{\Delta_{k}-\Delta_{j}}g_{2}(z)\ln(1-z)
\ee
where $g_{1}(z)$ is a rational function which has no singularities as $z \rightarrow \infty$, but which may have singularities as
$z\rightarrow 1$, while $g_{2}(z)$ is a polynomial in $1/z$ of rank
$\Delta_{i}-\tilde{\Delta}-1$. From this we see that the two operators
$\phi_{{i}}(Z,X,T)$ and $ {\cal O}_{{j}}(0,0)$ will not
commute once $X^2-T^2<0$.  That is, they will not commute when they are timelike separated on the boundary, even though
they are spacelike separated in the bulk. The
nonvanishing commutator comes only from the $\ln(1-z)$ term and is
thus proportional to $Z^{\Delta_{k}-\Delta_{j}}g_{2}(z)$.

One can define a new bulk operator
\begin{equation}
\phi_i(Z,X,T)=\int K_{\Delta_i}(Z,X,T|X',T')\mathcal{O}_i(X',T')+\sum_{n}d_{n}\int K_{\Delta_{n}}(Z,X,T|X',T')\mathcal{O}_{{n}}(X',T')
\label{extension}
\end{equation}
where $\Delta_{n}$ is an even number.  Given the structure of
the commutator we found above, each term in the sum contributes a polynomial in $\frac{1}{z} = \frac{T^2-X^2}{Z^2}$ of some rank.
One can adjust the coefficients $d_n$ in such a way as to cancel the commutator up to any desired power of
$1/z$. The problem with bulk locality arises when the points are timelike separated on the boundary but spacelike separated
in the bulk.  This corresponds to $|1/z|<1$. So canceling the commutator to a high power in $1/z$ means the
commutator can be made very small, except near the bulk
lightcone.  Depending on the operator content, it may even be possible to cancel the commutator to all orders
in $1/z$.

One might worry that this is all special to operators of even
conformal dimension, but this is not the case.  For non-integer
conformal dimensions (as arises for non-protected operators) the
appropriate analytic continuation (that is, analytic continuation of ${}_2F_{1}(\alpha,\alpha,\gamma,z)$
to $|z|>1$) gives
\begin{eqnarray}
&&\hspace*{-1cm}f(Z,X,T,0,0)=\frac{\Gamma(\Delta_{i})Z^{\Delta_{k}-\Delta_{j}}}{\Gamma(\tilde{\Delta})\Gamma(\Delta_{i}-\tilde{\Delta})}
\times \sum_{l=0}^{\infty} \frac{(\tilde{\Delta})_{l}(1+(\Delta_{j}-\Delta_{k}-\Delta_{i})/2)_{l}}{(l!)^2} \left(\frac{(T-i\epsilon)^2-X^2}{Z^2}\right)^{l}\nonumber \\
&&\qquad\times \left(2\psi(l+1)+\log(\frac{Z^2}{X^2-(T-i\epsilon)^2})-\psi(\Delta_{i}-\tilde{\Delta}-l)-\psi(\tilde{\Delta}+l)\right)
\end{eqnarray}
Here $\psi(x)=\frac{\Gamma '(x)}{\Gamma (x)}$ and
$(n)_{l}=\frac{\Gamma(n+l)}{\Gamma(n)}$ .  Again the log term gives
rise to a non-zero commutator when $X^2-T^2<0$, i.e.\ timelike separation on the boundary,
even if the points are spacelike separated in the bulk. The commutator has an expansion in
$\frac{(T-i\epsilon)^2-X^2}{Z^2}$ which for bulk
spacelike separation is less that $1$. Thus the structure is such that
by using (\ref{extension}) with appropriate $d_{n}$'s one can make the
commutator arbitrarily small, provided appropriate higher dimension operators exist.
Of course being able to carry out this procedure simultaneously for different pairs of operators
$\phi_i,\,{\cal O}_j$ will place stringent constraints on the operator content and interactions
of the CFT.

We have thus found that by adding higher dimension operators we can define local observables in the bulk.
In appendix \ref{sect:General-Mixed-Bulk/Boundary} we reach the same conclusion by smearing 3-point
correlators in the CFT.

\section{Bulk construction\label{sect:bulk}}

So far our approach has been to work purely within the CFT, seeking to
define bulk observables which commute at spacelike separation.  But let's
imagine that, at least in some approximation, we have access to a
local description of bulk physics.  Then we should be able to
re-derive our results from the bulk point of view.  Here we show how this
works, using AdS${}_2$ as our main example.

\subsection{Bulk equations of motion}

To illustrate how this works, take a massless $\phi^3$ theory in the bulk.
\be
\label{BulkAction}
S = \int d^2x \sqrt{-g} \left(-{1 \over 2} g^{\mu\nu} \partial_\mu \phi \partial_\nu \phi - {1 \over 3} \lambda \phi^3\right)
\ee
The bulk field is dual to an operator ${\cal O}$ with dimension 1 on the boundary.
The bulk equation of motion $\nabla \phi = \lambda \phi^2$ can be solved perturbatively in $\lambda$.
\[
\phi = \phi^{(0)} + \phi^{(1)} + \phi^{(2)} + \cdots
\]
where
\begin{eqnarray*}
&& \nabla \phi^{(0)} = 0 \\
&& \nabla \phi^{(1)} = \lambda \big(\phi^{(0)}\big)^2 \\
&& \nabla \phi^{(2)} = 2 \lambda \phi^{(0)} \phi^{(1)} \\
&& \qquad \vdots
\end{eqnarray*}
We already know how to solve the $0^{th}$ order equation.
\[
\phi^{(0)}(T,Z) = {1 \over 2} \int_{T - Z}^{T+Z} dT_1 \, {\cal O}(T_1)
\]
This can be represented diagrammatically as

\begin{center}
\includegraphics[width = 3.5cm]{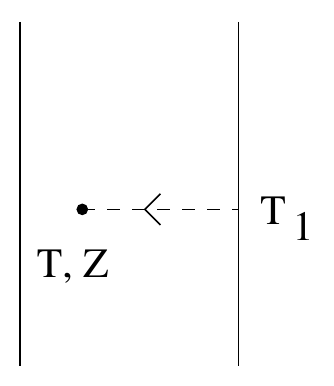}
\end{center}

\noindent
The dashed propagator is non-zero and equal to $1/2$ only in the right
lightcone of the bulk point $(T,Z)$.  The arrow on the dashed propagator
points towards the vertex of the lightcone.

The $1^{st}$ order equation is solved by
\[
\phi^{(1)}(x) = \int d^2 x' \sqrt{-g} \, G(x \vert x') \, \lambda \big(\phi^{(0)}(x')\big)^2
\]
where a suitable Green's function is
\[
G(T,Z \vert T',Z') = {1 \over 2} \theta(Z - Z') \theta(Z - Z' - \vert T - T' \vert)
\]
(non-zero and equal to $1/2$ only in the right light-cone of $(T,Z)$).
In defining the composite operator $\big(\phi^{(0)}\big)^2$ there is a
self-contraction one can make.  This generates a tadpole diagram that we will ignore.
More precisely, we have in mind canceling the tadpole against a linear term in the
action.  Dropping the tadpole amounts to normal-ordering $\big(\phi^{(0)}\big)^2$, so
\[
\phi^{(1)}(x) = {\lambda \over 2} \int_{\raisebox{-5pt}{\hbox{\rm right l.c. of $x$}}} \hspace{-1.5cm} d^2 x' \sqrt{-g} \,\,
\colon \big(\phi^{(0)}(x')\big)^2 \colon
\]
Writing this out explicitly
\[
\phi^{(1)}(T,Z) = {\lambda R^2 \over 8} \int_0^Z {dZ' \over (Z')^2} \int_{T-(Z-Z')}^{T+(Z-Z')} dT'
\int _{T'-Z'}^{T'+Z'} dT_1 dT_2 \, \colon {\cal O}(T_1) {\cal O}(T_2) \colon
\]
This is the first order correction introduced in (\ref{FirstOrder}).
By construction it's AdS covariant and satisfies the bulk equation of motion to
first order in $\lambda$.  It can be represented diagrammatically as

\begin{center}
\includegraphics[width = 3.5cm]{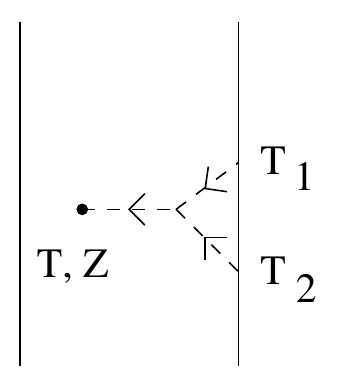}
\end{center}

In this diagram we're using the dashed propagator, and the vertex factor for three dashed lines
is $\lambda R^2 / (Z')^2$.

Likewise the 2${}^{nd}$ order equation is solved by
\be
\label{TriLocal}
\phi^{(2)}(x) = 2 \lambda \int d^2 x' \sqrt{-g} \, G(x \vert x') \, \phi^{(0)}(x') \phi^{(1)}(x')
\ee
which can be represented diagrammatically as

\begin{center}
\includegraphics[width = 4.375cm]{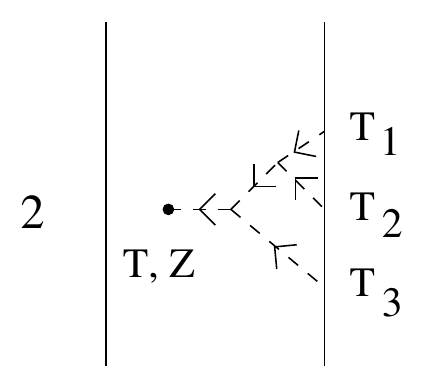}
\end{center}

There's an important difference between the procedure we have outlined here and conventional perturbation theory.
In conventional perturbation theory one begins with a free field that is local and causal and uses it as a basis
for building up an interacting field.  Superficially our construction is similar: we use $\phi^{(0)}$ as a basis for
constructing an interacting local bulk operator.  But note that, although $\phi^{(0)}$ obeys a free wave equation, it
is not a local field when interactions are taken into account in the CFT: as shown in section \ref{sect:problem} $\phi^{(0)}$
fails to commute with itself at spacelike separation.

\subsection{Bulk Feynman diagrams\label{sect:Feynman}}

In this section we show how the Feynman diagrams associated with a local theory in the bulk
can be mapped over to CFT calculations.  This will provide yet another way of deriving the CFT operators which are
dual to local bulk observables.  It will also show that, in a $1/N$ expansion of the CFT, these operators have
correlation functions which reproduce bulk perturbation theory.  As in the previous section, we work with
massless $\phi^3$ theory in the bulk as described by (\ref{BulkAction}).

We begin with a lemma.  From (\ref{Bulk2pt}) the bulk Feynman propagator is
\bea
\nonumber
i G_F(x \vert x') &=& \langle 0 \vert T\lbrace\phi(x) \phi(x')\rbrace \vert 0 \rangle \\
\nonumber
&=& {1 \over 2\pi} \tanh^{-1}
\left({2 Z Z' \over Z^2 + Z'{}^2 - (T - T')^2 + i \epsilon}\right) \\
\label{BulkFeynman}
&=& {1 \over 4 \pi} \, \log {(Z + Z')^2 - (T-T')^2 + i \epsilon \over (Z - Z')^2 - (T - T')^2 + i \epsilon}
\eea
Sending $Z \rightarrow 0$ gives the bulk-boundary Feynman propagator
\be
i G_F(T \vert x') = \langle 0 \vert T\lbrace{\cal O}(T) \phi(x')\rbrace \vert 0 \rangle = {Z' \over \pi} \,
{1 \over Z'{}^2 - (T - T')^2 + i \epsilon}
\ee
Consider applying the linear smearing relation (\ref{LowestOrder}) to the boundary operator ${\cal O}(T)$ which
appears here, in an attempt to recover the bulk Feynman propagator.  This gives
\be
{1 \over 2} \int_{T-Z}^{T+Z} dT_1 \, iG_F(T_1 \vert x') = {1 \over 4 \pi} \, \log {(Z + Z' + i \epsilon)^2 - (T-T')^2
\over (Z - Z' - i \epsilon)^2 - (T - T')^2}
\ee
Compared to the bulk Feynman propagator (\ref{BulkFeynman}), this has a different $i \epsilon$ prescription.
So -- unlike the Wightman functions considered in section \ref{sect:lowest} -- smearing the bulk-boundary Feynman
propagator does not give the bulk-bulk Feynman propagator.  Instead we find that the two expressions differ
in the right lightcone of the bulk point $(T,Z)$:
\be
\label{relation}
i G_F(x \vert x') = \int dT_1 \, K(x \vert T_1,0) \, iG_F(T_1 \vert x') + i \, K(x \vert x')\,.
\ee
Here
\be
K(T,Z \vert T',Z') = {1 \over 2} \, \theta(Z - Z') \theta\big(Z - Z' - \vert T - T' \vert\big)
\ee
is non-zero and equal to $1/2$ only when $(T',Z')$ lies in the right lightcone of the point $(T,Z)$.
Note that $K$ is exactly the Green's function we introduced in section \ref{sect:bulk}!  So (\ref{relation})
can be represented diagrammatically as

\begin{center}
\includegraphics{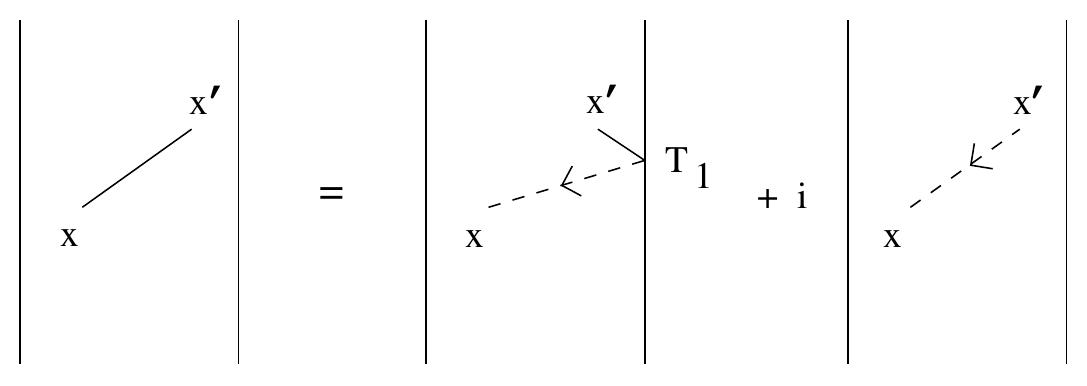}
\end{center}

\noindent
Here solid lines represent Feynman propagators $i G_F$ and dashed lines represent $K$.  This is the lemma we wished to
prove.

With this lemma it's straightforward to map bulk Feynman diagrams to CFT calculations.  For instance consider the
lowest-order Feynman diagram which contributes to $\langle \phi(T_1,Z_1) {\cal O}(T_2) {\cal O}(T_3) \rangle$.
Assuming
\be
\label{LCcond}
T_1 - Z_1 > T_2 > T_3
\ee
so that the operators are time-ordered and their right lightcones don't overlap
on the boundary, we have\footnote{Note that the vertex factor for three solid lines is $-i2\lambda R^2/Z^2$
while the vertex factor for three dashed lines is $\lambda R^2 / Z^2$.}
\be
\label{3ptfig}
\raisebox{-1.8cm}{\includegraphics{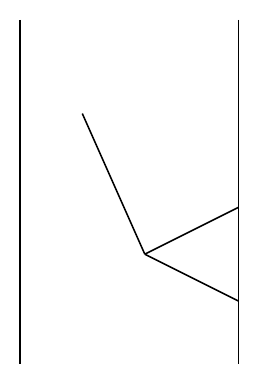}} = \raisebox{-1.8cm}{\includegraphics{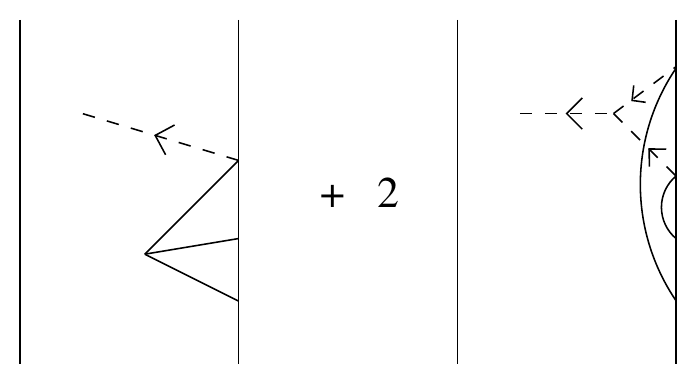}}
\ee
(We have dropped some diagrams involving dashed propagators from the bulk point to the boundary points.
They do not contribute since, given (\ref{LCcond}), the boundary points are not in the right lightcone of the
bulk point.)
The first diagram on the right hand side of (\ref{3ptfig}) involves
the CFT 3-point function, as induced by a bulk Feynman diagram (see
appendix \ref{sect:3point}).  The second diagram on the right
involves a disconnected product of CFT 2-point functions.  In terms of
correlators (\ref{3ptfig}) means
\be
\langle \phi(T_1,Z_1) {\cal O}(T_2) {\cal O}(T_3) \rangle =
\langle \phi^{(0)}(T_1,Z_1) {\cal O}(T_2) {\cal O}(T_3) \rangle +
\langle \phi^{(1)}(T_1,Z_1) {\cal O}(T_2) {\cal O}(T_3) \rangle
\ee
In other words, at this order computing $\langle (\phi^{(0)} + \phi^{(1)}) {\cal O} {\cal O} \rangle$
in the CFT exactly reproduces the tree-level correlator between one bulk point and two boundary points.
Moreover, from the last diagram in (\ref{3ptfig}) you can read off the need to include $\phi^{(1)}$
in the definition of a bulk observable.

As a more involved example, consider the correlator between two bulk points and one boundary point,
$\langle \phi(T_1,Z_1) {\cal O}(T_2) \phi(T_3,Z_3) \rangle$.  Taking $T_1 - Z_1 > T_2 > T_3 + Z_3$, so that
again the points are time-ordered and their right lightcones don't overlap, we have
\bea
\nonumber
\hspace*{-1cm}\raisebox{-2cm}{\includegraphics{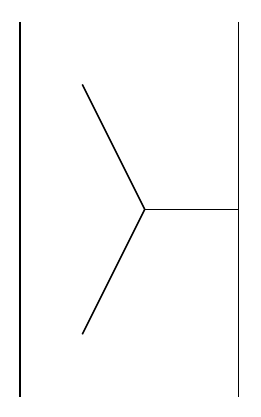}} & = & \raisebox{-2cm}{\includegraphics{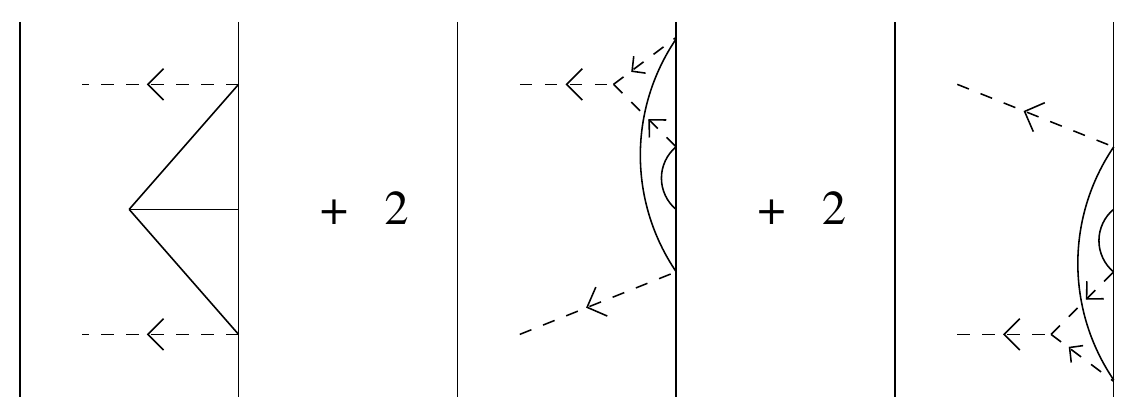}} \\[10pt]
\nonumber
& = & \! \langle \phi^{(0)}(x_1) {\cal O}(T_2) \phi^{(0)}(x_3) \rangle \,\, + \,\,
\langle \phi^{(1)}(x_1) {\cal O}(T_2) \phi^{(0)}(x_3) \rangle \,\, + \,\,
\langle \phi^{(0)}(x_1) {\cal O}(T_2) \phi^{(1)}(x_3) \rangle \\
\eea
In the first diagram on the right the lowest order smearing functions
are tied together with a three-point correlator in the CFT.  In the
second and third diagrams the first order correction to the smearing
function is combined with a disconnected product of CFT two-point
correlators.  So again we see that to this order in $\lambda$ we can
identify the combination $\phi^{(0)} + \phi^{(1)}$ with a local operator in the bulk.

\section{Generalizations and extensions\label{sect:extensions}}

In this section we discuss the extension of our results, first to
general CFT's, then to higher orders in $1/N$, using AdS${}_2$ to
illustrate the ideas.

Consider a general one-dimensional CFT, with primary fields ${\cal O}_i$ of dimension $\Delta_i$.  The
3-point correlator is a generalization of (\ref{Bdy3pt}).
\bea
&&\langle 0 \vert {\cal O}_i(T_1) {\cal O}_j(T_2) {\cal O}_k(T_3) \vert 0 \rangle \\
\nonumber
&& = c_{ijk} \,
{1 \over (T_1 - T_2)^{\Delta_i + \Delta_j - \Delta_k} \, (T_1 - T_3)^{\Delta_i + \Delta_k - \Delta_j
} \,
(T_2 - T_3)^{\Delta_j + \Delta_k - \Delta_i}}
\eea
The simplest way to construct CFT operators dual to bulk observables is to generalize the construction of
section \ref{sect:bulk} and note that this correlator is induced at tree level by a cubic coupling between
bulk scalar fields.  The bulk action is a generalization of (\ref{BulkAction}),
\be
S = \int d^2x \sqrt{-g} \left(-{1 \over 2} g^{\mu\nu} \partial_\mu \phi_i \partial_\nu \phi_i
- {1 \over 2} m_i^2 \phi_i^2 - {1 \over 3} \lambda_{ijk} \phi_i \phi_j \phi_k \right)
\ee
where $\Delta_i = {1 \over 2} \left(1 + \sqrt{1 + 4 m_i^2 R^2}\right)$, and the coefficient
of proportionality relating $c_{ijk}$ and $\lambda_{ijk}$ could be worked out as in appendix
\ref{sect:3point}.
At lowest order we have the expression for bulk observables worked out in section 3.1 of \cite{Hamilton:2005ju},
\be
\label{GeneralPhi0}
\phi_i^{(0)}(x) = \int dT' \, K_{\Delta_i}(x \vert T') {\cal O}_i(T')
\ee
where the smearing function for an operator of dimension $\Delta$ is
\be
K_\Delta(T,Z \vert T') = {\Gamma(\Delta + 1/2) \over \sqrt{\pi} \Gamma(\Delta)} \left({Z^2 - (T-T')^2 \over Z}\right)^{\Delta-1} \theta\left(Z - \vert T - T' \vert\right)
\ee
These lowest-order operators satisfy a free equation of motion, $\big(\nabla - m_i^2\big) \phi_i^{(0)} = 0$.
The first order correction, satisfying $\big(\nabla - m_i^2\big) \phi_i^{(1)} = \lambda_{ijk} \phi_j^{(0)}
\phi_k^{(0)}$, is given by
\be
\label{GeneralPhi1}
\phi_i^{(1)}(x) = \lambda_{ijk} \int d^2x' \sqrt{-g} \, G_{\Delta_i}(x \vert x') \, \phi_j^{(0)}(x') \phi_k^{(0)}(x')
\ee
where an appropriate Green's function, satisfying $\big(\nabla - m^2\big) G_\Delta(x \vert x') = {1 \over \sqrt{-g}}
\delta^2(x - x')$, is
\be
G_\Delta(x \vert x') = {1 \over 2} P_{\Delta-1}(\sigma) \theta(Z - Z') \theta(Z - Z' - \vert T - T' \vert)\,.
\ee
This Green's function was worked out in section 2.2 of \cite{Hamilton:2005ju}.  It is non-zero only in the right lightcone
of the point $x$.  $\sigma$ is the invariant distance (\ref{InvDistance}) between $x$ and $x'$, and $P_{\Delta-1}$ is
a Legendre function.

By construction the operators $\phi_i^{(0)} + \phi_i^{(1)}$ satisfy the bulk equations of motion
to first order in $\lambda$.  They will commute at spacelike separation, along the lines of section
\ref{sect:bilocal}, provided that 4-point and higher-point correlators are ignored in the CFT.  Thus
(\ref{GeneralPhi0}) and (\ref{GeneralPhi1}) define a local bulk observable in any one-dimensional CFT,
to the extent that 4 and higher point correlators can be ignored.

A natural conjecture is that this pattern continues order-by-order when higher-point correlators are taken into account.
For instance, to build commuting bulk observables when 4-point correlators are taken into account, we should add
a correction $\phi_i^{(2)}$ which is tri-local in the CFT primaries.  For an explicit example of a second-order
correction, in the CFT dual to $\phi^3$ theory in the bulk, see (\ref{TriLocal}).

This conjecture is consistent with leading large-$N$ counting.  Recall that in the
't Hooft large-$N$ limit the connected correlation function of $k$ single-trace operators scales as
\[
\langle {\cal O}_1 \cdots {\cal O}_k \rangle_C \sim 1/N^{k-2}
\]
In the $\phi^3$ theory of section \ref{sect:bulk}, tree diagrams with $k$ external legs scale as $\lambda^{k-2}$, so we can
identify the bulk coupling $\lambda \sim 1/N$.  The idea is that a bulk observable has an expansion
\be
\phi = \sum_{n = 0}^\infty \phi^{(n)}
\ee
where $\phi^{(n)}$ carries an explicit factor of $\lambda^n$ and is a multi-local expression involving $n+1$ single-trace
operators, defined so that there are no self-contractions.  To fix $\phi^{(n+1)}$ the recipe is as follows.  Suppose we have
already constructed $\phi^{(0)},\ldots,\phi^{(n)}$
so that bulk operators commute at the level of $(n+2)$-point functions.  Taking the connected $(n+3)$-point
correlator into account will lead to a non-zero commutator in
\be
\label{problem}
\left\langle \big[ \phi^{(n)}, {\cal O} \big] {\cal O} \right\rangle \sim \lambda^{n} {1 \over N^{n+1}}
\sim {1 \over N^{2n + 1}}
\ee
There's no reason to expect this to vanish, so we need to further correct our definition of a
bulk observable.  We conjecture that $\phi^{(n+1)}$ can be chosen to cancel (\ref{problem}), at least at
spacelike separation.  As a consistency check, at least the powers of $N$ come out the same:
\be
\left\langle \big[ \phi^{n+1}, {\cal O} \big] {\cal O} \right\rangle \sim \sum_k \lambda^{n+1} {1 \over N^{k-2}} {1 \over N^{n+2-k}}
\sim {1 \over N^{2n + 1}}
\ee
(the CFT correlator is a sum of disconnected products of $k$-point and $(n+4-k)$-point correlators).

This argument shows that the conjecture is consistent with planar large-$N$ counting.  Of course there are subleading
non-planar corrections to CFT correlators: recall that a CFT diagram with $k$ punctures and $L$ handles, dual to a bulk
diagram with $k$ legs and $L$ loops, scales as $1/N^{2L+k-2}$.  It should be possible to take these non-planar corrections
into account by making subleading corrections to the bulk operators.

To get a feel for the sort of corrections which arise from non-planar diagrams, consider the following diagram in the
$\phi^3$ theory of section \ref{sect:bulk}.

\begin{center}
\includegraphics[width = 3.5cm]{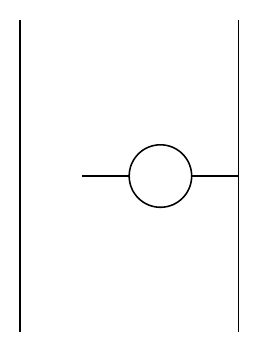}
\end{center}

\noindent
From the bulk point of view this diagram is ${\cal O}(\lambda^2)$; it
is a one-loop correction to the bulk-boundary propagator.  The form of
the bulk-boundary propagator is fixed by AdS invariance, so this
diagram can be absorbed into a mass and wavefunction renormalization
of the bulk field.  From the CFT point of view this diagram is an
${\cal O}(1/N^2)$ effect.  Mapping it to the CFT as in section
\ref{sect:Feynman} gives\footnote{The bulk Feynman diagram has a
symmetry factor of $1/2$, which we write out explicitly in the
coefficients of the diagrams with a dashed propagator.}

\begin{center}
\hspace*{-1cm}\includegraphics[width=15cm]{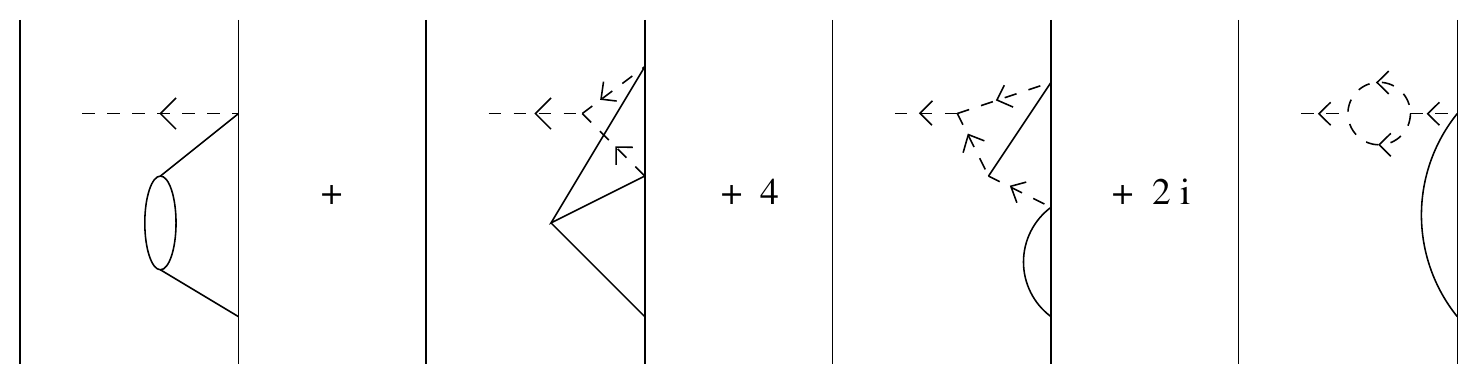}
\end{center}

\noindent
A few comments are in order.
\begin{itemize}
\item
The first diagram is a contribution to $\langle \phi^{(0)} {\cal O} \rangle$.  It makes an
${\cal O}(1/N^2)$ correction to the CFT 2-point function, correcting the conformal dimension of the boundary operator.\footnote{Bulk perturbation theory corresponds to
a large-$N$ expansion of CFT correlators.  This expansion in powers of $\lambda$ shouldn't be confused with
(\ref{Bdy3pt}), where $\lambda R^2$ was defined as the coefficient of the exact CFT 3-point function.}
\item
The second diagram is a contribution to $\langle \phi^{(1)} {\cal O} \rangle$, involving the CFT 3-point function
evaluated at ${\cal O}(1/N)$.  Note that the CFT correlator is induced by a conventional bulk Feynman diagram,
which means it's a time-ordered product.  This is the first place where operator ordering is important, and it suggests
that the CFT operators appearing in (\ref{FirstOrder}) should be time-ordered.
\item
The final two diagrams involve a 2-point function in the CFT.  They can be thought of as making an
${\cal O}(1/N^2)$ correction to the
lowest-order smearing function (\ref{LowestOrder}), appropriate for the corrected conformal dimension coming from the first diagram.
\end{itemize}

\section{Conclusions\label{sect:conclusions}}

In this paper we described the construction of local bulk operators
from the CFT beyond leading order in $1/N$. This provides a working
definition of Heisenberg bulk operators, which may be used to
construct new off-shell bulk quantum gravity amplitudes directly from
conformal field theory correlators.  We showed that using the naive
smeared operator beyond the leading large-$N$ limit results in bulk
operators which do not commute at spacelike separation. We then
showed that this problem can be cured in perturbation theory, by
changing the definition of bulk operators.  We presented several
derivations of the corrected operators.  The most interesting
constructions -- adding higher-dimension operators, and adding
multi-local corrections -- could be carried out completely within the
CFT.  In these constructions one seems to need a large number of
primary operators with prescribed properties to make the bulk theory
local. The requisite properties seem to be satisfied in a large $N$
CFT, to all orders in $1/N$ perturbation theory, through the presence of
multi-trace operators with appropriate insertions of derivatives.  An alternate
construction uses the radial Hamiltonian from a local bulk theory in
AdS.\footnote{A method for defining bulk operators, based on identifying the radial
Hamiltonian with the Fokker-Planck Hamiltonian of the boundary theory, was developed in \cite{Lifschytz:2000bj}.}
The different constructions agree in perturbation theory, but the
CFT construction may make it possible to understand how bulk locality
breaks down.

One might be surprised that such a construction could be carried out at all,
since the diffeomorphism constraints of quantum gravity would seem to
rule out the existence of local observables
\cite{Kuchar:1991qf,Isham:1992ms}.  Of course we constructed our
observables purely within the CFT, where we never really had to face
up to this issue: normalizeable diffeomorphisms in the bulk act
trivially on the CFT.  This means our bulk observables are by
construction diffeomorphism invariant.  This suggests that, from the
bulk point of view, we have managed to construct local observables
using a particular choice of gauge (corresponding to our use of Poincar\'e
coordinates to label points in the bulk).

To better understand these results let us look at a local quantum
field theory on AdS (without gravity). Then the limit of bulk
correlation functions as you approach the boundary still look like
those of a CFT.  The bulk operators (which are all independent at
some fixed time) can be written as integrals over the boundary operators
(at different times) using the radial Hamiltonian approach. This gives
local bulk operators, but this is not a surprise since there really is
a local bulk theory and we have just exchanged the initial data
surface (at fixed time) with an initial data surface on the timelike
boundary. In this case one can either use the radial Hamiltonian
approach or regular perturbation theory around local free fields, both
should give the same answer.  The key difference between the
perturbative expansion using the radial Hamiltonian which we used in
the previous sections and the usual perturbative expansion of Green
functions in quantum field theory, is that beyond leading order in the perturbation
expansion the operator $\phi^{(0)}$ which we used does not commute with
itself at bulk spacelike separation. However $\phi^{(0)}$ does satisfy the free wave equation,
so for a given bulk Lagrangian, it will produce correlators that agree
order by order in the coupling with correlators constructed in the
usual interaction picture approach. This expansion may be viewed as a
choice of non-local interpolating field being used to set up the
perturbative expansion. Since the operators we construct satisfy the
correct operator equation of motion, they provide the same
approximation to the full Heisenberg operator as standard perturbation theory.
It's important to note that in this setup the
boundary theory is not unitary.  Things can come in from the bulk, since
there really are extra degrees of freedom in the bulk not accounted
for on the boundary at some fixed time.  Nevertheless, boundary
correlation functions do look like those of a CFT.

Now add gravity. If we just do perturbation theory to some order in
$1/N$ around an AdS background things work as above (as long
as the theory is renormalizable). The bulk theory is local, the
boundary operators look like a sector in a CFT, and writing bulk
operators using the radial Hamiltonian will of course give local bulk
operators.

However in the full quantum gravity theory (meaning finite $N$, and not perturbation around a
fixed background to some order) things are different. The only local
operators are at the boundary, which means there are fewer degrees of
freedom.  This is manifested by the fact that the boundary theory is now unitary.  A unitary
theory on the boundary cannot describe a local QFT in the bulk.

From the CFT point of view the most plausible way for bulk locality to
fail is if the constraints on the CFT primaries, that we needed to construct
local bulk observables in section \ref{sect:ads3}, cannot be satisfied beyond some conformal dimension $\Delta_{\rm max}$.  Let's say $\Delta_{\rm max}$ is of order $N$.
What are the consequences of this?  The infinite sum over primaries (\ref{extension})
that is necessary for locality is
truncated, so bulk operators will not commute at spacelike separation.
Take a bulk operator at a point $(Z,X=0,T=0)$ and a
boundary operator at $(X=0,T)$. The commutator of the two operators
inside a correlation function, as long as all other operators are far
away, is $[\phi(Z,0,0),{\cal O}(0,T)]
\sim\frac{1}{N}(\frac{T^2}{Z^2})^{N}$. This means the
causal structure of the bulk spacetime has been destroyed. However away
from the bulk lightcone the commutator is of order $e^{-N}$, which is invisible in
perturbation theory. Very near the lightcone, say $Z^2-T^2 \sim
(1-\frac{a}{N})$ with $a$ independent of $N$, the commutator will
be non-zero even in perturbation theory. But the
interpretation, in perturbation theory, is just that one has a
slightly non-local bulk theory, with non-locality on the scale of $1/N$.
These represent the expected light-cone fluctuations, and not complete destruction of bulk
space-time locality, even though non-perturbatively the whole bulk causal
structure is destroyed. When other operators are nearby the condition
for non-commutativity changes somewhat. For a three point
function the condition is $\chi<1$ where $\chi$ is given in
(\ref{eq:crossratio}).

How far one can venture from the lightcone and still see a large commutator?
The answer depends on $Z$.  It is of order $\delta T \sim aZ/N$,
or $\delta X \sim Z \sqrt{a/N}$. So for very large $Z$ (i.e.\ near
the Poincar\'e horizon) there is a large region on the boundary, which is
space-like to a bulk point, and in which operators will have a large
commutator with the bulk point. This is due to the large redshift from
the boundary to the Poincar\'e horizon. This is just the old argument
about small non-locality near the horizon getting transmitted to large scales
on the boundary and giving rise to a
stretched horizon.

Finally, it is worth trying to draw conclusions from these results
regarding generic predictions which might be used to motivate future
experimental tests of theories of quantum gravity.  An important observation is that to all orders in the $1/N$ expansion we have
constructed local bulk observables whose $n$-point correlators respect both causality and AdS covariance.
At finite $N$ presumably causality is violated, along the lines discussed above, but exact
AdS covariance is maintained.
Other models of quantum gravity predict modified dispersion relations
arising from violation of Lorentz invariance on short distance scales
\cite{AmelinoCamelia:1997gz}.
Using AdS covariance as a proxy for Lorentz invariance, the present
work predicts there should be no sign of such modified dispersion
relations. This is compatible with recent experimental results
\cite{Ackermann:2009zq} which bound the scale of such corrections at well above
the Planck scale.
Instead the results of the present
work indicate that new quantum gravity effects are only to be
expected once one looks for signs of causality violation -- operators that fail to commute at spacelike separation --
in 3- and higher-point functions.

\bigskip
\centerline{\bf Acknowledgements}
We thank the Aspen Center for Physics where this work was initiated.
DK is supported by U.S.\ National Science Foundation grant PHY-0855582
and PSC-CUNY awards 60038-39-40 and 63332-00-41.  He is grateful to
the 2010 Simons Workshop for hospitality during the course of this
research and would like to thank Debajyoti Sarkar and Shubho Roy for
valuable discussions.  The research of GL is supported in part by the
Israeli Science Foundation under grant no.\ 392/09. GL would also like
to thank the Erwin Schr\"odinger Institute for hospitality while this
research was in progress.  The research of D.A.L.\ is supported in part
by DOE grant DE-FG02-91ER40688-Task A. D.A.L.\ thanks the Harvard
Physics Department for hospitality while part of this work was
completed.

\appendix
\section{Cubic couplings in AdS${}_2$\label{sect:3point}}

The $\phi^3$ interaction of section \ref{sect:bulk} induces a tree-level 3-point coupling
\bea
\nonumber
&& \langle 0 \vert T\lbrace \phi(x_1) \phi(x_2) \phi(x_3) \rbrace \vert 0 \rangle = - {i \lambda R^2 \over 4 \pi^3}
\int_{-\infty}^\infty dT_4 \int_0^\infty {dZ_4 \over Z_4^2} \, \tanh^{-1} \left({1 \over \sigma_{14} + i \epsilon}\right) \\
\label{Bulk3pt}
&& \qquad \tanh^{-1} \left({1 \over \sigma_{24} + i \epsilon}\right) \tanh^{-1} \left({1 \over \sigma_{34} + i \epsilon}\right)
\eea
Sending the bulk points to the boundary gives
\be
\label{Feynman3pt}
\langle 0 \vert T\lbrace {\cal O}(T_1) {\cal O}(T_2) {\cal O}(T_3) \rbrace \vert 0 \rangle = - {i \lambda R^2 \over \pi} \,
{1 \over \vert T_1 - T_2 \vert \cdot \vert T_1 - T_3 \vert \cdot \vert T_2 - T_3 \vert}
\ee
This agrees with (\ref{Bdy3pt}) for $T_1 > T_2 > T_3$.  So we can think of our 3-point coupling in the CFT
as coming from a bulk Feynman diagram

\begin{center}
\includegraphics[width = 3.5cm]{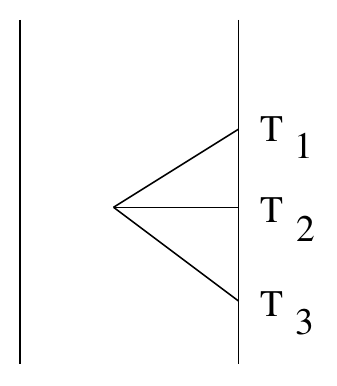}
\end{center}

\noindent
Since the form of the 3-point function is fixed by conformal invariance, the only real
lesson here is that our bulk and boundary conventions for normalizing $\lambda$ are
the same.

\section{Commutators at large $N$\label{appendix:HDO}}

We wish to calculate
\[
\langle 0 \vert i \big[\phi^{(1)}_{\Delta = 2}(T,Z), {\cal O}(T_2)\big] {\cal O}(T_3) \vert 0 \rangle
\]
where
\[
\phi^{(1)}_{\Delta = 2}(T,Z) = A \int_{T-Z}^{T+Z} dT' \,\, {Z^2 - (T-T')^2 \over Z} \,
\colon{\cal O}(T') {\cal O}(T')\colon
\]
We begin by studying the correlator
\[
\langle 0 \vert \phi^{(1)}_{\Delta = 2}(T,Z) {\cal O}(T_2) {\cal O}(T_3) \vert 0 \rangle
\]
with $T - Z > T_2 > T_3$ so the operators don't overlap.  At leading order for large $N$ the correlator
is given by a disconnected product of two CFT 2-point functions (\ref{Wightman}).  This gives
\beas
&& \langle 0 \vert \phi^{(1)}_{\Delta = 2}(T,Z) {\cal O}(T_2) {\cal O}(T_3) \vert 0 \rangle \\
&& = {2 A \over \pi^2} \int_{T-Z}^{T+Z} dT' \,\, {Z^2 - (T - T')^2 \over Z} \, {1 \over (T'-T_2)^2} \, {1 \over (T' - T_3)^2} \\
&& = - {8 A \over \pi^2 (T_2 - T_3)^2} + {4 A (Z^2 - T^2 + T T_2 + T T_3 - T_2 T_3 \over \pi^2 Z (T_2 - T_3)^3}
\log {(T+Z-T_3)(T-Z-T_2) \over (T-Z-T3)(T+Z-T_2)}
\eeas
This can be continued into the regime $T+Z > T_2 > T-Z > T_3$ with a $T_2 \rightarrow T_2 + i \epsilon$ prescription.
We then repeat the calculation, starting from
\[
\langle 0 \vert {\cal O}(T_2) \phi^{(1)}_{\Delta = 2}(T,Z) {\cal O}(T_3) \vert 0 \rangle
\]
with $T_2 > T+Z$.  Continuing to the same regime as before gives exactly the same expression, but with a $T_2 \rightarrow
T_2 - i \epsilon$ prescription.  Taking the difference gives the commutator
\[
\langle 0 \vert i \big[\phi^{(1)}_{\Delta = 2}(T,Z), {\cal O}(T_2)\big] {\cal O}(T_3) \vert 0 \rangle
= {8 A \over \pi} \, {1 \over (T_2 - T_3)^2} \, \psi
\]
where the AdS invariant cross-ratio $\psi$ is defined in (\ref{psi}).
The calculation of 
\[
\langle 0 \vert i \big[\phi^{(1)}_{\Delta = 4}(T,Z), {\cal O}(T_2)\big] {\cal O}(T_3) \vert 0 \rangle
\]
proceeds along the same lines.

\section{Mixed bulk-boundary correlators and conformal invariance\label{sect:boundary-to-bulk-map}}

We work in Poincar\'e coordinates where the three-dimensional AdS
metric takes the form\[
ds^{2}=\left(dZ^{2}+dX^{2}-dT^{2}\right)/Z^{2}\]
The isometries of this metric form the group $SO(2,2)$ which is generated
by the following symmetry transformations: $SO(1,1)$ Lorentz transformations
on $x^{\mu}=(T,X)$; dilatations, acting as\[
Z\to\lambda Z,\qquad x^{\mu}\to\lambda x^{\mu}\]
and special conformal transformations, parametrized by $b^{\mu}$,
acting as 
\begin{eqnarray}
x^{\mu} & \to & \frac{x^{\mu}-b^{\mu}(x^{2}+Z^{2})}{1-2b\cdot x+b^{2}(x^{2}+Z^{2})}\nonumber \\
Z & \to & \frac{Z}{1-2b\cdot x+b^{2}(x^{2}+Z^{2})}\,.\label{eq:specialcon}
\end{eqnarray}
The {}``bulk'' distance function transforms as
\[
|x_{1}-x_{2}|^{2}+Z_{1}^{2}+Z_{2}^{2}\to\frac{|x_{1}-x_{2}|^{2}+Z_{1}^{2}+Z_{2}^{2}}{\left(1-2b\cdot x_{1}+b^{2}(x_{1}^{2}+Z_{1}^{2})\right)\left(1-2b\cdot x_{2}+b^{2}(x_{2}^{2}+Z_{2}^{2})\right)}\,.\]
In the limit that $Z\to0$ these expressions reduce to the familiar
global conformal transformations of two-dimensional conformal field
theory. In the following, it will be helpful to define $\gamma_{x,z}=1-2b \cdot x+b^{2}(x^{2}+Z^{2})$.

Let $\mathcal{O}(X,T)$ be a CFT primary operator with conformal dimension
$\Delta$. The dual bulk scalar operator according to the prescription
of \cite{Hamilton:2006az} is 
\begin{eqnarray}
\phi(Z,X,T)&=&\int dx'dt'K_{\Delta}(Z,X,T|X',T')\mathcal{O}(X+ix',T+t')\\
\nonumber
&=&\frac{\Delta-1}{\pi}\int_{x'^{2}+t'^{2}<Z^{2}}dx'dt'\left(\frac{Z^{2}-x'^{2}-t'^{2}}{Z}\right)^{\Delta-2}\mathcal{O}(X+ix',T+t')\label{eq:smear}
\end{eqnarray}
Correlators of this operator with other CFT primary operators transform
covariantly under the group $SO(2,2)$. To see this consider acting
with such a transformation on the mixed bulk-boundary correlator
\begin{equation}
\hspace*{-2cm}
\left\langle \phi(Z,X,T)\prod_{k}\mathcal{O}_{k}(x_{k}^{\mu})\right\rangle =\frac{\Delta-1}{\pi}\int_{x'^{2}+t'^{2}<Z^{2}}dx'dt'\left(\frac{Z^{2}-x'^{2}-t'^{2}}{Z}\right)^{\Delta-2}\left\langle \mathcal{O}(X+ix',T+t')\prod_{k}\mathcal{O}_{k}(x_{k}^{\mu})\right\rangle \label{eq:mixedcorel}\end{equation}
The expression is manifestly dilatation covariant, and Lorentz invariant,
so it remains to check special conformal transformations \eqref{eq:specialcon}.
The CFT correlator transforms covariantly under such a transformation.
We wish to check whether
\begin{equation}
\left\langle \phi(\tilde{Z},\tilde{X},\tilde{T})\prod_{k}\mathcal{O}_{k}(\tilde{x}_{k}^{\mu})\right\rangle =\left\langle \phi(Z,X,T)\prod_{k}\mathcal{O}_{k}(x_{k}^{\mu})\right\rangle \prod_{j}\gamma_{x_{j},0}^{\Delta_{j}}\label{eq:tobechecked}
\end{equation}
where $\tilde{Z}$, etc. are related to $Z$, etc. via the transformation
\eqref{eq:specialcon}. Using \eqref{eq:mixedcorel} the left-hand
side of \eqref{eq:tobechecked} is\begin{equation}
\hspace*{-1cm}\left\langle \phi(\tilde{Z},\tilde{X},\tilde{T})\prod_{k}\mathcal{O}_{k}(\tilde{x}_{k}^{\mu})\right\rangle =\prod_{k}\gamma_{x_{k},0}^{\Delta_{k}}\int_{a^{2}+b^{2}<\tilde{Z}^{2}}dadb\left(\frac{\tilde{Z}^{2}-a^{2}-b^{2}}{\tilde{Z}}\right)^{\Delta-2}\gamma_{Y,0}^{\Delta}\left\langle \mathcal{O}(Y^{\mu})\prod_{k}\mathcal{O}_{k}(x_{k}^{\mu})\right\rangle \label{eq:transform}\end{equation}
using covariance of the CFT correlator, and defining $A=(ia,b)$ which
are related to new dummy variables $x'',y''$ and $Y^{\mu}=(X+ix'',T+t'')$
by a special conformal transformation\begin{equation}
\left(\tilde{x}+A\right)^{\mu}=\frac{Y^{\mu}-b^{\mu}Y^{2}}{1-2b\cdot Y+b^{2}Y^{2}}\label{eq:changeofvars}\end{equation}
Now\[
\frac{\tilde{Z}^{2}-a^{2}-b^{2}}{\tilde{Z}}=\frac{1}{\gamma_{Y,0}}\frac{Z^{2}-x''^{2}-t''^{2}}{Z}\]
and\[
dadb=\frac{1}{\gamma_{Y,0}^{2}}dx''dt''\]
We therefore find that the $\gamma_{Y,0}$ factors in the integrand
of \eqref{eq:transform} cancel. However one must bear in mind that
the change of variables \eqref{eq:changeofvars} makes $x''$ and
$t''$ complex, though the surface of integration is still bounded
by the locus $Z^{2}-x''^{2}-t''^{2}=0$. For infinitesimal transformations,
it is clear the integral can be viewed as a double contour integration,
and each contour can be deformed back to the disc, where $x''$ and
$t''$ are real. Therefore we can finally switch dummy variables and
recover \eqref{eq:tobechecked}. Thus we conclude the mixed bulk-boundary
correlator \eqref{eq:mixedcorel} transforms covariantly under $SO(2,2)$.

\section{General bulk / boundary three-point function\label{sect:General-Mixed-Bulk/Boundary}}

Let us consider the mixed bulk/boundary three-point function with
a bulk operator (dual to operator of conformal weight $\Delta$) and
two boundary operators with conformal weights $\Delta_{1}$ and $\Delta_{2}$.
We use the results of appendix \ref{sect:boundary-to-bulk-map}
to first express the general functional form of the correlator. A
key point to note is that a cross-ratio, invariant under dilatations
and special conformal transformations, can be constructed using two
boundary points and one bulk point
\begin{equation}
\chi=\frac{\left(\left(\vec{x}-\vec{x}_{1}\right)^{2}+Z^{2}\right)\left(\left(\vec{x}-\vec{x}_{2}\right)^{2}+Z^{2}\right)}{Z^{2}\left(\vec{x}_{2}-\vec{x}_{1}\right)^{2}}\label{eq:crossratio}
\end{equation}
Let us define 
\begin{equation}
\left \langle \phi_{\Delta}(Z,\vec{x}){\cal O}_{\Delta_{1}}(\vec{x}_{1}){\cal O}_{\Delta_{2}}(\vec{x}_{2})\right \rangle=c(Z,\vec{x};\vec{x}_{1};\vec{x}_{2})
\end{equation}
Using dilatations, rotations, translations and special conformal transformations,
the general three-point function may be fixed to be of the form
\begin{equation}
\hspace*{-1cm}c(Z,\vec{x};\vec{x}_{1};\vec{x}_{2})=|\vec{x}_{1}-\vec{x}_{2}|^{-(\Delta_{1}+\Delta_{2}-\Delta)}\left[\frac{Z^{2}+(\vec{x}-\vec{x}_{1})^{2}}{Z}\right]^{-(\Delta+\Delta_{1}-\Delta_{2})/2}\left[\frac{Z^{2}+(\vec{x}-\vec{x}_{2})^{2}}{Z}\right]^{-(\Delta+\Delta_{2}-\Delta_{1})/2}f(\chi)
\end{equation}

The form of the function $f(\chi)$ may be fixed by performing a conformal
transformation to send point $\vec{x}_{1}\to0$ and $\vec{x}_{2}\to\infty$
and comparing to the results of section \ref{sect:ads3}.
In this limit \[
\chi\to\frac{|\vec{x}|^{2}+Z^{2}}{Z^{2}}\]
so\begin{equation}
c(Z,\vec{x};0;\infty)=Z^{\Delta}\left(|\vec{x}|^{2}+Z^{2}\right)^{-(\Delta+\Delta_{1}-\Delta_{2})/2}f\left(\frac{|\vec{x}|^{2}+Z^{2}}{Z^{2}}\right)\label{eq:threeptinf}\end{equation}
which should be matched with \eqref{eq:threeptcoeff}. This fixes
\[
f(\chi)=\frac{1}{2\pi R}\left(\frac{\chi}{\chi-1}\right)^{(\Delta+\Delta_{1}-\Delta_{2})/2}\,_{2}F_{1}\left(\left(\Delta+\Delta_{1}-\Delta_{2}\right)/2,\:\left(\Delta+\Delta_{1}-\Delta_{2}\right)/2;\:\Delta;\:\frac{1}{1-\chi}\right)\]
It is a nontrivial fact that this expression is symmetric under switching
$1\leftrightarrow2$. This completes the derivation of the general
three-point function.

The singularities of the three-point function occur at $\chi=0,1$.
The locus $\chi=0$ coincides with the bulk light-like separations
between the bulk point and one of the boundary points. However the
locus $\chi=1$ yields singularities at bulk spacelike separations
in general. In the special limit when $ $one of the boundary points
moves off to infinity, this simply becomes the boundary light-cone.
More generally the position of the singularity is sensitive to the
position of both boundary operators.

There are three interesting limits that \eqref{eq:threeptinf} may
be expanded around. The basic CFT limit is extracted from the $Z^{\Delta}$
coefficient in the limit that $Z\to0$ with $\vec{x}$ fixed. In this
limit $f\to1/2\pi R$ so the expected CFT behavior of $|\vec{x}|^{-(\Delta+\Delta_{1}-\Delta_{2})}$
is recovered.

The OPE of the gravitational theory is recovered by expanding around
the bulk light-cone $\chi=0$. This yields an expression of the form
\begin{equation}
c(Z,\vec{x};0;\infty)\sim c_{1}Z^{\Delta_{2}-\Delta_{1}}\left(1+\mathcal{O}(\chi)\right)+c_{2}Z^{\Delta_{2}-\Delta_{1}}\chi^{\Delta_{2}-\Delta_{1}}\left(1+\mathcal{O}(\chi)\right)\label{eq:bulkope}
\end{equation}
where $c_{i}$ are constants. The first term is analytic in $\chi$,
and hence respects bulk causality. The second term can lead to noncommutativity,
but only at timelike bulk separations. Overall, the bulk OPE is of
the form expected from Wilson's original paper \cite{Wilson:1969zs},
namely an expansion in a function of the bulk geodesic distance.

The problem we need to attend to  comes from examining the correlator \eqref{eq:threeptinf}
around the locus $\chi=1$. When point 2 is at infinity, this corresponds
to boundary light-like separations of points 0 and 1. More generally,
this locus simply corresponds to bulk spacelike or timelike separations,
depending on the position of operator 2. Expanding around this locus
we find\[
c(Z,\vec{x};0;\infty)\sim Z^{\Delta_{2}-\Delta_{1}}c_{4}\log\left(\chi-1\right)\left(1+\mathcal{O}(\chi-1)\right)+c_{3}\left(1+\mathcal{O}(\chi-1)\right)\]
The presence of the log term leads to non-commutativity at bulk spacelike
separations. But as in section \ref{sect:ads3} it can be canceled by adding higher dimension operators.


\providecommand{\href}[2]{#2}\begingroup\raggedright\endgroup

\end{document}